\documentclass[traditabstract]{aa} 
\usepackage{graphicx}
\usepackage{natbib}

\usepackage{txfonts}
\newcommand{\Hw}{Hawk-I }
\newcommand{\U}{UDS }
\newcommand{\G}{GOODS-S }

\begin{document}
  \title{The Hawk-I UDS and GOODS Survey (HUGS): Survey design and
    deep K-band number counts }


  \author{A. Fontana
   \inst{1}
    \and J. S. Dunlop \inst{2}
    \and D. Paris \inst{1}
    \and T. A. Targett \inst{2,3}
    \and K. Boutsia \inst{1}
   \and M. Castellano \inst{1}
  \and A. Galametz \inst{1}
    \and A. Grazian \inst{1}
  \and R. McLure \inst{2}
    \and E. Merlin \inst{1}
    \and L. Pentericci \inst{1}
   \and S. Wuyts \inst{4}
   \and O. Almaini \inst{5}
  \and K. Caputi \inst{6}
   \and R.-R. Chary \inst{7}
  \and M. Cirasuolo \inst{2}
 \and C. J. Conselice  \inst{5}
   \and A. Cooray \inst{8}
   \and E. Daddi \inst{9}
   \and M. Dickinson \inst{10}
    \and S. M. Faber \inst{11}
   \and G. Fazio \inst{12} 
   \and H. C. Ferguson \inst{13}
   \and E. Giallongo \inst{1}
   \and M. Giavalisco \inst{14}
 \and N. A. Grogin \inst{13}
   \and N. Hathi \inst{15}
   \and A. M. Koekemoer \inst{13}
   \and D. C. Koo \inst{11} 
   \and R. A. Lucas \inst{13}
 \and M. Nonino \inst{16}
   \and H. W. Rix \inst{17}
    \and A. Renzini \inst{18}
    \and D. Rosario \inst{4}
    \and P. Santini \inst{1}
    \and C. Scarlata \inst{19}
    \and V. Sommariva \inst{1,21}
   \and D. P. Stark \inst{20}      
    \and A. van der Wel \inst{17}
    \and E. Vanzella \inst{21}
    \and V. Wild \inst{22,2}
\and H. Yan \inst{23}
   \and S. Zibetti \inst{24} 
     }
    \institute{INAF - Osservatorio Astronomico di Roma - Via Frascati 33 Monte Porzio Catone 00040 Rome Italy
      \email{adriano.fontana@oa-roma.inaf.it}
      \and
    Institute for Astronomy, University of Edinburgh, Royal Observatory, Edinburgh EH9 3HJ, UK 
      \and 
  Department of Physics and Astronomy, Sonoma State University, Rohnert Park, CA, SA
      \and
    Max-Planck-Institut f¨ur extraterrestrische Physik (MPE), Giessenbachstrasse 1, D-85748, Garching bei Munchen, Germany  
     \and 
    The School of Physics and Astronomy, University of Nottingham, Nottingham, UK 
      \and
 Kapteyn Astronomical Institute, University of Groningen, PO Box 800, NL-9700 AV Groningen, the Netherlands
      \and
    Spitzer Science Center, MC 220–6, California Institute of Technology, 1200 East California Boulevard, Pasadena, CA 91125, USA
      \and
    Department of Physics and Astronomy, University of California, Irvine, CA, USA 
     \and 
   CEA-Saclay/DSM/DAPNIA/Service d’Astrophysique, Gif-sur-Yvette, France  
      \and
    National Optical Astronomy Observatories, Tucson, AZ, USA  
     \and 
    UCO/Lick Observatory, Department of Astronomy and Astrophysics,
    University of California, Santa Cruz, CA, 95064, USA
     \and 
    Harvard-Smithsonian Center for Astrophysics, Cambridge, MA, USA  
     \and 
    Space Telescope Science Institute, Baltimore, MD, USA  
     \and 
  Department of Astronomy, University of Massachusetts, Amherst, MA, USA
      \and
 Aix Marseille Universit\'e, CNRS, LAM (Laboratoire d'Astrophysique de Marseille) UMR 7326, 13388, Marseille, France
      \and
    INAF - Osservatorio Astronomico di Trieste, Trieste, Italy  
      \and
    Max Planck Institute for Astronomy, Konigstuhl 17, 69117 Heidelberg, Germany 
      \and 
    Osservatorio Astronomico di Padova, Padova, Italy
      \and
    Minnesota Institute of Astrophysics and School of Physics and Astronomy, University of Minnesota, Minneapolis, MN, USA
     \and 
    Department of Astronomy, Steward Observatory, University of Arizona, Tucson, AZ, USA  
     \and  
   INAF Osservatorio Astronomico di Bologna, Bologna, Italy   
     \and
School of Physics and Astronomy, University of St Andrews, North Haugh, St Andrews, KY16 9SS
  \and 
    Department of Physics and Astronomy, University of Missouri, Columbia, MO, USA
    \and 
    INAF-Osservatorio Astrofisico di Arcetri, Largo Enrico Fermi 5, 50125 Firenze, Italy
   }

  \date{Received ; accepted}

 
  \abstract{We present the results of a new, ultra-deep, near-infrared 
    imaging survey executed with the Hawk-I imager at the ESO VLT, of which we make all the data (images and catalog) public. This survey, named 
    HUGS (Hawk-I UDS and GOODS Survey), provides deep, high-quality 
    imaging in the $K$ and $Y$ bands over the portions of the UKIDSS UDS and GOODS-South fields 
    covered by the CANDELS HST WFC3/IR survey. 
    In this paper we 
    describe the survey strategy, the observational campaign, the
    data reduction process, and the data quality. We show that, thanks to exquisite image
    quality and extremely long exposure times, HUGS delivers the
    deepest $K$-band images ever collected over areas of cosmological
    interest, and in general ideally complements the CANDELS data set in
    terms of image quality and depth.  In the GOODS-S field, the
    $K$-band observations cover the whole CANDELS area with a complex
    geometry made of 6 different, partly overlapping pointings, in order to
    best match the deep and wide areas of CANDELS imaging.
    In the deepest region 
    (which includes most of the {\it Hubble} Ultra Deep Field) exposure times exceed 
    80 hours of integration, yielding a
    $1-\sigma$ magnitude limit per square arcsec of $\simeq 28.0$\,{\it AB mag}. 
    The seeing is exceptional and homogeneous 
    across the various pointings, confined to the range $0.38 - 0.43$\,arcsec.
    In the UDS field the survey is about one magnitude shallower (to match the
    correspondingly shallower depth of the CANDELS images) but includes
    also $Y$-band band imaging (which, in the UDS, was not provided 
    by the CANDELS WFC3/IR imaging). In the $K$-band, with an
    average exposure time of 13 hours, and seeing in the range 
    $0.37 - 0.43$\,arcsec, the $1-\sigma$ limit per square
    arcsec in the UDS imaging is $\simeq 27.3${\it AB mag}.  In the $Y$-band,  
    with an average exposure time $\simeq 8$\,hours, and seeing in the range
    $0.45 - 0.5$\,arcsec, the imaging yields a $1-\sigma$ limit per square
    arcsec of $\simeq 28.3${\it AB mag}. 
    We show that the HUGS
    observations are well matched to the depth of the CANDELS WFC3/IR data, since
    the majority of even the faintest galaxies detected in the CANDELS $H$-band images
    are also detected in HUGS.  Finally we present the $K$-band galaxy number counts 
    produced by combining the HUGS data from the two fields.
    We show that the slope of the number counts depends
    sensitively on the assumed distribution of galaxy sizes, with
    potential impact on the estimated extra-galactic background light.}
 
 \keywords{surveys; galaxies: evolution}
 \maketitle
 %
\section{Introduction}

Ultra-deep imaging surveys are of fundamental importance for advancing our knowledge 
of the early phases of galaxy formation and
evolution. In general, each technological advance in telescopes and/or detectors 
has been swiftly applied to obtain ever deeper images of the extragalactic sky 
over a range of wavelengths. At optical wavelengths the use of the first CCDs 
to obtain early galaxy number counts (\citet{Ellis1997} and references therein) 
provides an obvious example of this technology-driven progress, as do the early 
{\it Hubble} Deep Field (HDF) campaigns \citep{Williams1996,Williams2000}, which paved the way 
for the subsequent exploration of the high-redshift Universe. Many ground-based
telescopes have been exploited in this way since the end of the
last century, with nearly every new instrumental set-up available being used 
to obtain deep observations within various survey fields
(e.g. the NTT Deep Field \citep{Fontana2000}, the Keck Deep Field \citep{Savicki2005}, the
VLT Fors Deep Field (\citep{Heidt2003} etc).  

In recent years, the emphasis has started to shift progressively towards 
undertaking deep imaging surveys in the near-infrared. 
The most recent and spectacular case is the long
series of {\it Hubble} Ultra Deep Field (HUDF) campaigns (\cite{Illingworth2013} and references therein), 
obtained with the Wide Field Camera 3 (WFC3/IR), the latest and most efficient 
instrument on board {\it HST}. The observations secured in the various bands 
from the  $Y_{098}$ to the $H_{160}$ represent our deepest view of the Universe, reaching a final
depth that in some case exceeds the 31$^{st}$ magnitude 
(e.g. \cite{Ellis2013,Koekemoer2013}.

The shift to near-infrared surveys has not only been driven by technological 
advances, but is motivated by the need to sample the rest-frame optical (and even UV) 
emission in highly-redshifted galaxies in the young Universe.

In this context, deep ground-based $K$-band surveys remain of fundamental importance even in the WFC3 era.  
It is worth remembering that, at $z = 6$, the wavelength gap between $H$ and 3.6\,$\mu$m 
is comparable to the gap between the observed $Z$ and $K$ bands at $z = 3$. 
Thus, bridging this large spectral range with deep $K$-band photometry 
is crucial for an accurate determination of the rest-frame physical quantities 
(e.g. stellar age, stellar mass, dust content) of galaxies at very high redshifts. 
As an example, the wavelength shift from the $H_{160}$ band (the longest accessible from {\it HST}) 
to the $K$ band enables us to extend the redshift coverage of the rest-frame $B$ band from $z \simeq 2.6$ to 
$z \simeq 4$. In addition, at $z > 3.5$, imaging longward of the $H$-band is needed to locate and 
measure the size of the Balmer break, which reaches IRAC 3.6\,$\mu$m only at $z > 8$. 
The $K$-band is also considered an excellent proxy for the selection of mass-selected samples of 
galaxies at high redshift: at $z < 2$, {\it Spitzer} imaging is not really required for the derivation of 
stellar masses \citep{Fontana2006}, and even at higher redshifts the depth and image quality 
typically obtained with ground-based $K$-band imaging makes $K$-band surveys competitive 
with the deepest {\it Spitzer} data sets. For these reasons, most of the newly-introduced near-infrared 
imagers have been used to secure progressively deeper fields in the $K$-band 
\citep{Elston1988,Cowie1990,Moustakas1997,Huang1997,Cristobal-Hornillos2003,Labbe2003, Minowa2005,Grazian2006,Caputi2006,Conselice2008,Retzlaff2010, McCracken2010,  Cirasuolo2010, McCracken2012}.

Needless to say, an ultra-deep field in a single bandpass is of
relatively little scientific use in isolation. For this reason, most of the
surveys mentioned above have been targetted on a few, carefully-selected, 
high galactic latitude fields, in order to accumulate deep multi-wavelength exposures
across as many bandpasses as possible. Building on the experience of
the early HDF, the concepts of colour selection criteria and photometric
redshifts have become a common tool in the exploration of galaxies at high redshifts.

The CANDELS survey (PI S. Faber, Co-PI H. Ferguson) is the latest, and most ambitious 
enterprise of this kind.  As described in \cite{Grogin2011, Koekemoer2011}, CANDELS 
is a 900-orbit {\it HST} multi-cycle treasury  program delivering 0.18-arcsec $J$ (F125W) and $H$ 
(F160W) images reaching 27.2 (AB mag; 5$\sigma$) over 0.25 sq. degrees, with even deeper ($\simeq 28$\,AB mag; 5$\sigma$) 
3-band ($Y$, $J$, $H$) images over $\simeq 120$\, sq. arcmin (within GOODS-South and GOODS-North). 
It also delivers the necessary deep optical ACS parallels to complement the deep WFC3/IR imaging. 
The major scientific goals of this program are the assembly of statistically-useful 
samples of galaxies at $6 < z < 9$, measurement of the morphology and internal colour structure of 
galaxies at $z = 2 - 3$, the detection and follow-up of SuperNovae at $z > 2$ for validating 
their use as cosmological distance indicators, and the study of the growth of black-holes 
in the centres of high-redshift galaxies. With deep {\it Spitzer} imaging
data \citep{Ashby2013} and ultra-deep radio imaging available in all 
five fields, deep X-ray imaging either available or planned, and {\it Herschel}/Laboca 
imaging (plus ongoing JCMT/SCUBA2 imaging) also now provided at sub-mm wavelengths, 
the legacy value of the {\it Spitzer} SEDS and {\it HST} CANDELS data is clear and unrivalled.

While optical ground-based or ACS imaging is available over most of
the CANDELS fields, the $K$-band coverage is generally patchy and inadequate.
Because of the small field-of-view of the ISAAC imaging on VLT, 
the depth of the pre-existing VLT ISAAC $K$-band mosaic over GOODS-S \citep{Retzlaff2010}
is significantly shallower than required to match the new WFC3/IR observations
from CANDELS, and is of inhomogeneous quality. To fill this important 
gap in the available multi-wavelength coverage, we therefore
designed a survey that makes optimum use of the latest near-infrared imager
on the VLT. This instrument, 
the High Acuity Wide field K-band Imager (Hawk--I, \cite{Kissler2008}) delivers 
high-quality imaging over a relatively large field-of-view, delivering square
images 7.5\,arcmin on a side, with exquisite sampling of even the best 
point-spread-function (PSF) delivered by the VLT (the pixel size is 0.11\,arcsec) 
and state-of-the-art quantum efficiency and detector cosmetics. 
Being optimally matched to the size of the CANDELS fields,
Hawk-I can provide an unprecedented combination of depth and area
coverage. Our survey targets two of the three CANDELS fields accessible from
Paranal, namely GOODS-S and the UDS, since the ongoing UltraVISTA survey is already
delivering ultra-deep $Y$, $J$, $H$, $K$ imaging within the COSMOS CANDELS 
field \citep{McCracken2012}. We have named this survey 
HUGS (Hawk-I UDS and GOODS Survey)
to emphasize the unique role of Hawk-I in enabling the delivery 
of this important data set.

Although HUGS is designed to complement the WFC3 CANDELS observations,
it will also enable scientific investigations on its own, thanks to
the depth of the $K$-band images. Among these, we mention for instance
the analysis of the spectral energy distribution of $z\simeq 4$
galaxies, where the $K$-band allows us to sample with high accuracy the
Balmer break of selected LBGs (see Castellano et al. 2014) or
the evolution of the galaxy mass function at high redshift (see Grazian
et al., subm.), where $K$-band selected samples are required to
be as complete in mass as possible.

This paper describes (and the accompanying data release provides) a
complete compilation of the data obtained within HUGS incorporating also 
all the previous observations that we executed with Hawk-I on GOODS-S 
in previous campaigns; the GOODS-S field was already observed with Hawk-I in the
framework of the Hawk-I Science Verification and subsequently in a previous
ESO Large Program aimed at identifying $z\simeq 7$ galaxies.  These programs
have delivered a robust estimate of the Luminosity Function
of $z\simeq 7$ galaxies (\cite{Castellano2010a,Castellano2010b}), and led to the
discovery of the first robust spectroscopic confirmation at $z> 7$
(\cite{Vanzella2011}).

In this paper we describe the survey design and the data collected from 
the completed survey.  We provide accurate estimates of the final quality
of the data, in terms of both depth and image quality; the latter is
particularly impressive, given the long integrations from the ground
that have been used. Finally, we use these new data to obtain the deepest
galaxy number counts ever secured in the $K$-band over a statistically
meaningful area.  We have used AB magnitudes throughout \citep{Oke1974}.

\section{Survey strategy}

The HUGS survey was designed to cover the two CANDELS fields
accessible from Paranal that do not
have suitably-deep $K$-band images: a sub-area of the UKIDSS Ultra Deep Survey 
(hereafter UDS) and GOODS-South (hereafter GOODS-S).

The depth of the images in the $K$-band has been tuned in order to match
the depth of the WFC3/IR images produced in the $J_{125}$ and $H_{160}$
filters. In practice, the target depth was chosen to be 0.5\,{\it mag} shallower
than obtained with WFC3/IR in $H_{160}$, as appropriate 
to match the average $H-K$ colour of faint galaxies.

In both fields deep $Y$-band images have also been acquired. In the case
of the UDS, these images provide an essential complement to 
the CANDELS data set, since neither
$Y_{098}$ nor $Y_{105}$ imaging of this field has been obtained within CANDELS. In the
case of GOODS-S, the  $Y$-band images come from an earlier program designed to
select $z\simeq 7$ galaxies with ground-based images
\cite{Castellano2010a,Castellano2010b}. These images are slightly less deep than the
$Y_{105}$ images that have since been obtained within CANDELS, and
cover only about 70\% of the GOODS-S field, but have nevertheless been 
reduced and are made available here as part of HUGS. 
We describe below the details of the two fields, in terms of pointings, exposure time and expected depth.

We note that, since \Hw is a mosaic of four
square detectors (2k$\times$2k each), it delivers images that exhibit a shallower crossed region at the
centre of the mosaic. Although our dithering pattern has been chosen
to minimise its impact, this feature is inevitable in the output data.

\subsection{The UDS field}

  \begin{figure}
  \centering
  \includegraphics[width=8.5truecm]{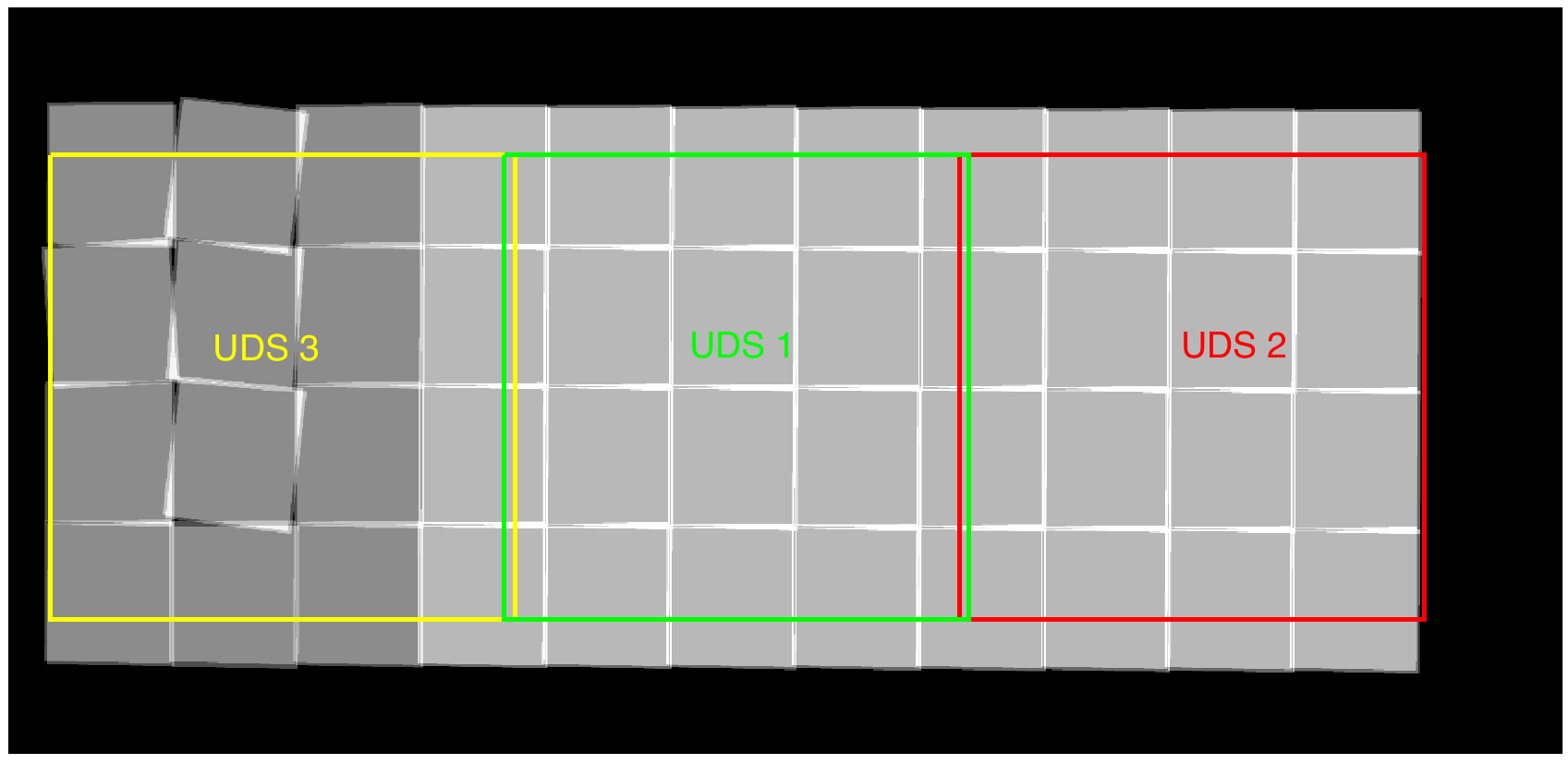}
  \caption{Location of the three Hawk-I pointings overlaid
    on the exposure map of the  WFC3-CANDELS  mosaic of the UDS. The greyscale of the WFC3/IR
    images is on a linear stretch from 0 to 4ks.
  }
     \label{UDSlayout}
  \end{figure}
Thanks to its quite regular shape, the UDS CANDELS field has been straightforward
to cover with Hawk-I. Three different \Hw pointings are able to cover
85\%  of the \U field. The layout is shown in
Figure\ref{UDSlayout}. We show the position of the three different
pointings (named UDS1, UDS2 and UDS3 in the following), assuming a
nominal size of 7.5 $\times$7.5 arcmin for the \Hw image. 
It can be seen that the three pointings
also provide two overlapping regions that have been used to
cross-check the photometric and astrometric solutions in the three
individual mosaics.  The three pointings have been exposed with
nearly identical exposure times, of 8 hours in the $Y$ band and 13
hours in the $K_s$ band (final exposure times are slightly different
since some images have been discarded during the reduction process).
Table~\ref{UDStable} summarises the location and exposure times of the various pointings.

\subsection{The GOODS-S field}

  \begin{figure}
  \centering
  \includegraphics[width=8.5truecm]{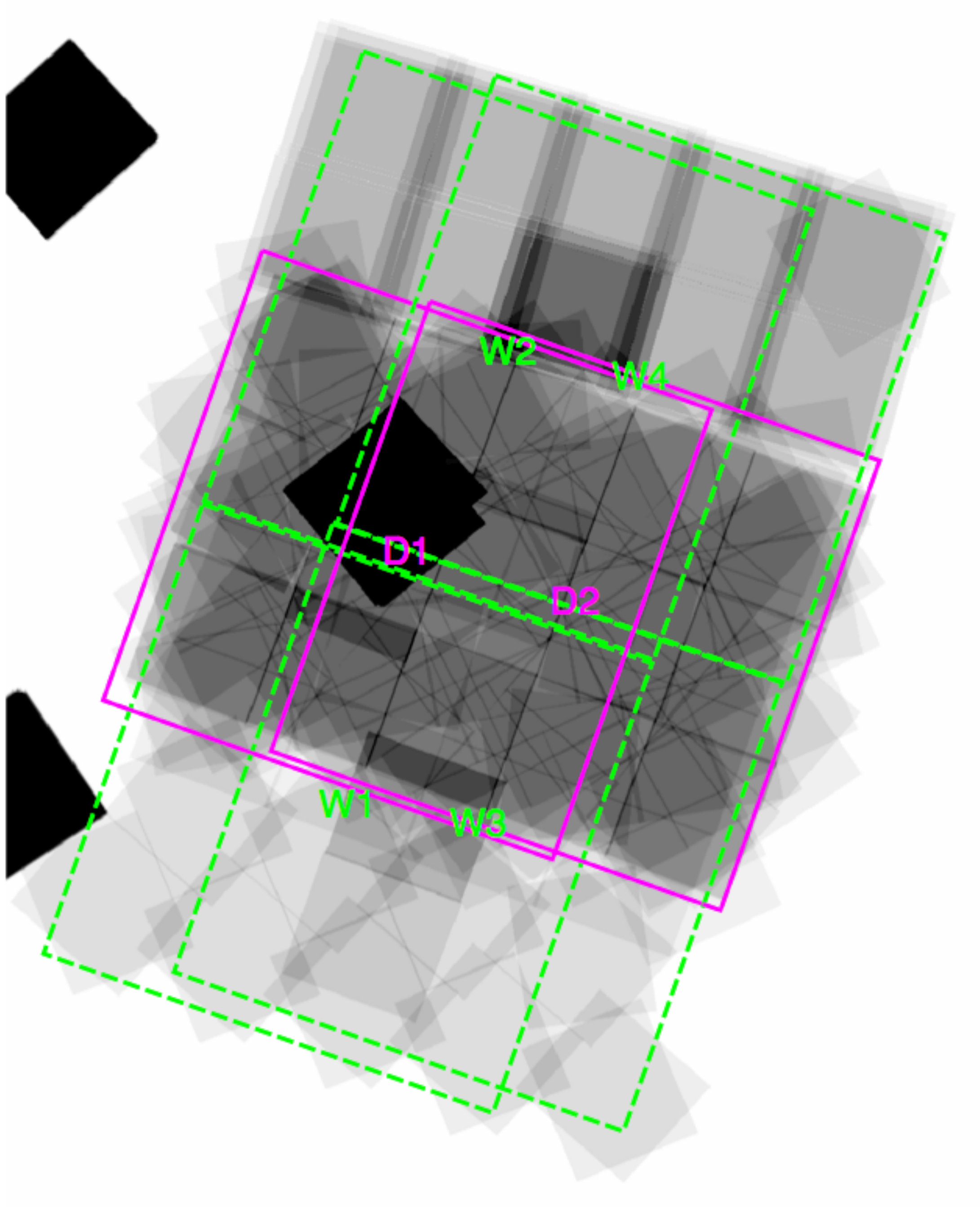}
  \caption{Location of the Hawk-I pointings overlaid
    on the exposure map of the WFC3 data available within the GOODS-South field. 
    Magenta lines show the field-of-view of pointings D1 and D2, while the 
    green lines show pointings W1, W2, W3 and W4 (see Table\ref{GOODStable}). 
    The black square at the centre is the HUDF12 region (Koekemoer et al. 2013). 
    The greyscale of the WFC3/IR
    images is on a linear stretch that saturates at the deepest levels
    of the CANDELS data; the HUDF12 is deeply oversaturated. 
       }
     \label{GOODSlayout}
  \end{figure}

  The coverage of the \G field with CANDELS is more complex, and
  forced us to adopt a more complicated pattern for the HUGS observations.
  The WFC3 observations are deeper in a rectangular region (10 by 7
  arcmin, named GOODS--Deep in CANDELS) that spans the entire
  width (i.e. east-west extent) of the \G field, and is centered in the
  vertical (i.e. north-south) direction. The 
  HUDF is located close to the centre of this area. These deep images
  are complemented by shallower WFC3/IR images (partly from the ERS
  survey \citep{Windhorst2011} and partly by CANDELS) that cover the 
  remaining area of the original ACS optical mosaic. 
  Our final layout has been designed to deliver a
  deeper image over GOODS-Deep, while also covering nearly the whole
  CANDELS area. Since the width of the \G field is 10\,arcmin, it cannot
  be covered efficiently with a single \Hw pointing. We therefore
  decided to cover the whole field with a $2 \times 3$ grid of
  pointings, rotated by $-19.5$ degrees to parallel the ACS and WFC3 mosaics.
  The layout is shown in Figure\ref{GOODSlayout}: as for the UDS, we
  show both the position of the 6 individual pointings overlaid upon
  the WFC3 exposure map (that includes also the position of the UDF
  and the other parallel deep fields) as well as the final 
  exposure map.

  The pointings are offset in the W/E direction by 3\,arcmin each, and
  in the N/S direction by 6\,arcmin each.  This approach has also made optimal use
  of the $K_s$-band images obtained in our earlier program. The two
  central pointings have been exposed for a total of about 31 hours, 
  while the four upper and lower pointings have been exposed for about 11 hours each. We
  therefore name GD1 and GD2 the two deep exposures, and GW1, GW2,
  GW3, GW4 the four shallower ones \footnote{For those readers who
    would dare downloading and reducing the raw data from scratch,
    these are named GOODS1 or GOODS-D1, GOODS-D2, GOODS-WIDE1,
    GOODS-WIDE2, GOODS-WIDE3 and GOODS-WIDE4 in the Observing Block description.}
  This layout also has the advantage of producing a final mosaic where
  each region of the GOODS-Deep area is observed with different
  physical regions of the instrument, further minimizing possible
  trends due to large-scale residuals in the flat-fielding.

Looking at  Figure~\ref{GOODSlayout} it is immediately appreciated that the coverage of 
the GOODS-Deep area is not uniform, because of the combined effects of the pointing 
locations and the \Hw gaps. It reaches nearly 90\,hours of exposure time in the very 
central area that covers most of the UDF, and still reaches more than 40\,hours of exposure 
time over the remaining part of the GOODS-Deep area. This image currently provides a unique 
combination of depth and area in the $K$-band.

During the earliest observations, several frames were acquired in the $H$ and $Br$-$\gamma$ filters, 
over the GD1 pointing. They were taken accidently due to a temporarily miscalibration of the 
Hawk-I filter wheel. We have also reduced these images, and make them available, 
although they have not been calibrated nor used in any scientific analysis. 

Table~\ref{GOODStable} summarises the location and exposure time of the various pointings.

\begin{table*}
\caption{Layout and summary of observations for the UDS field.}
\label{UDStable}
\centering
\begin{tabular}{c c c c c c c c  }   
\hline\hline
Pointing & Central RA & Central DEC  & Area (arcmin$^2$) &Exp. time
(s/hr) & Final seeing & maglim\tablefootmark{a} & maglim\tablefootmark{b} \\
\hline
\multicolumn{7}{c}{$K$-band}\\
\hline
UDS1 & 02:17:37.470 & $-$05:12:03.810 & 70 & 48360 / 13.43 & 0.37 & 27.4 & 26.1\\
UDS2 & 02:17:07.943& $-$05:12:03.810 & 70 &  46820 / 13.00  & 0.43 & 27.3 & 25.9\\
UDS3 & 02:18:06.896 &$-$05:12:03.810 & 70 & 45240 / 12.57 & 0.41 & 27.4 & 25.9\\
\hline
\multicolumn{7}{c}{$Y$-band}\\
\hline
UDS1 & 02:17:37.470 & $-$05:12:03.810 & 70 & 28800 / 8.00 & 0.45 & 28.4 & 26.9\\
UDS2 & 02:17:07.943& $-$05:12:03.810 & 70 & 28800 / 8.00 & 0.50 & 28.3 & 26.7\\
UDS3 & 02:18:06.896 & $-$05:12:03.810 & 70 & 29400 / 8.17 & 0.48 & 28.2 & 26.6\\
\hline
\multicolumn{7}{c}{$H$-band}\\
\hline
UDS1 & 02:17:37.470 & $-$05:12:03.810 & 70 & 13800 / 13.83 & 0.44 & N.A. & N.A.\\
\hline
\multicolumn{7}{c}{$Br$-$\gamma$-band}\\
\hline
UDS1 & 02:17:37.470 & $-$05:12:03.810 & 70 & 5760 / 1.6 & 0.41 & N.A. & N.A.\\
UDS2 & 02:17:07.943& $-$05:12:03.810 & 70 & 5760 / 1.6 & 0.42 & N.A. & N.A.\\
\hline
\end{tabular}

\tablefoot{
\tablefoottext{a}{at 1$\sigma$ arcsec$^{-2}$}
\tablefoottext{b}{at 5$\sigma$ in 1 FWHM} 
}

\end{table*}

\begin{table*}
\caption{Layout and summary of observations for the GOODS-S field. }
\label{GOODStable}
\centering
\begin{tabular}{c c c c c c c c  }   
\hline\hline
Pointing$^{(a)}$ & Central RA & Central DEC  & Area (arcmin$^2$) & Exposure time (s/hr) & Final seeing & maglim$^{(b)}$ & maglim$^{(c)}$ \\

\hline
\multicolumn{7}{c}{$K$-band}\\
\hline
GOODS-D1 & 03:32:36.835 & $-$27:47:45.24 & 70 & 113520 / 31.53 & 0.39 & 27.8 & 26.5\\
GOODS-D2 & 03:32:24.890 & $-$27:48:33.22 & 70 & 112800 31.33 & 0.38 & 27.8 & 26.5\\
GOODS-W1 & 03:32:41.080 & $-$27:51:44.32 & 70 & 47220 /13.12 & 0.43 & 27.4 & 26.0\\
GOODS-W2 & 03:32:29.650 & $-$27:44:37.26 & 70 & 40800 / 11.33& 0.38 & 27.3 & 26.0\\
GOODS-W3 & 03:32:31.796 & $-$27:52:01.74 & 70 & 37320 /10.37 & 0.38 & 27.3 & 25.9\\
GOODS-W4 & 03:32:20.242 & $-$27:44:59.97 & 70 & 41880 /11.63 & 0.42 & 27.3 & 25.8\\
\hline
\multicolumn{7}{c}{$H$-band}\\
\hline
GOODS-D1 & 03:32:36.835 & $-$27:47:45.24 & 70 & 21360 / 5.93 & 0.42 & N.A. & N.A.\\
\hline
\end{tabular}

\tablefoot{
\tablefoottext{a}{Each pointing has been rotated to PA=$-$19.5\,deg.}
\tablefoottext{b}{at 1$\sigma$ arcsec$^{-2}$}
\tablefoottext{c}{at 5$\sigma$ in 1 FWHM} 
}
\end{table*}

\section{Data acquisition and reduction}

\subsection{Observations}
All imaging in the $K$-band was obtained with individual images of 10 seconds of integration, 
averaged in sets of 12 images during acquisition (in ESO jargon these two parameters are 
referred to as DIT and NDIT, respectively). In the case of the $Y$-band images we 
adopted DIT=30 and NDIT=4.  A random dithering pattern with a typical offset of 12\,arcsec 
was applied in all cases. Observations were scheduled in Observing Blocks of about 1\,hr of 
execution each, corresponding to about 45 and 48 minutes of exposure in $K$ and $Y$, respectively. 
The position angle of each Observing Block (OB)  was rotated by 90\,degrees, so that the 
final mosaics are the results of individual images obtained with different physical regions 
of the detectors (which allowed us to test the accuracy of the photometric and calibration procedures).

All observations were executed in Service Mode, over a series of ESO observing periods from P86 to P90.
We retrieved from the archive all the images obtained during these runs, including those that 
were not graded within specifications during the observations. All these 
images were then analysed with an automated pipeline to assess their quality. 
We have included in the final coadded frames also some of the images graded ``out of specs'', 
but have excluded those with wildly-discrepant seeing, poor photometric quality  
or other cosmetic defects. For this reason the actual exposure times listed in 
Tables \ref{UDStable} and \ref{GOODStable} are slightly different from pointing to pointing, 
even though we originally planned identical exposure times.

\subsection{Data reduction}

We initially used two pipelines to independently reduce the images acquired in the first year 
of observations. One pipeline has been developed at the Rome Observatory, and is derived from 
a pipeline used to reduce LBT imaging data both in the visible and in the infrared. 
In its former version it was used to reduce the earliest \Hw data in GOODS-S (\cite{Castellano2010a,Castellano2010b}.
The second pipeline has been developed at the University of Edinburgh (originally 
to reduce UKIRT WFCAM and Gemini NIRI data; \cite{Targett2011}). We then
compared the two pipelines and the resulting reduced images, both in 
terms of conceptual steps and algorithms adopted, as well as in terms of 
the final image mosaic. The two results agreed very well and this comparison 
was utilised to help yield a final, optimised version of the Rome pipeline.
This refined Rome pipeline has been used in the final processing of all the data, 
including a re-processing of any previously-reduced images. 
For this reason the images produced here are slightly different/better than those 
used in \cite{Castellano2010b}. The final reduction pipeline is described in more 
detail in a separate paper (Paris et al in prep.) but we describe here 
the basic steps (that follow the usual recipe of infrared imaging 
data reduction) and specific features.

\subsubsection{Pre--reduction}
The raw images were retrieved from the ESO archive and each
Observing Block was processed separately at this stage.  The
initial reduction procedure consists of the removal of the dark
current and application of a flat-field in order to normalise the response
of each image pixel. Each flat--field is obtained from twilight
sky-flats
obtained in the same day of the observations, or in the day
immediately before or after when not available. Typically between 30
and 60 images are used to compute the flat--field. At first each scientific frame is subtracted by
a median stack dark image obtained by combining a set of dark frames
with the same EXPTIME and NDIT values of the observation set images.
Then a median stack flat image (masterflat) is created, by combining a
set of sky flats taken with the same filter as the observation set,
each subtracted by its own dark. While combining, each flat is
normalised by its own median background level, in order to obtain a
final median stack flat normalised to unity. Each scientific image is
then divided by the masterflat, so the response of pixels is finally
homogenised. During this pre-reduction stage pixel masks are also created to
flag saturated regions, cosmic ray events, satellite tracks and bad
(hot/cold) pixels. We have also developed a specific procedure to take
the effects of persistence into account - i.e. the residual counts left by
bright objects on the subsequent images. After some tests, we decided to identify all
the pixels that are above $10^4$ counts in a given image and mask them
out from the following image. Given the large number of subsequent
images, this efficiently masks out most of the pixels affected by
persistence. We assume that the remaining contribution is efficiently
wiped out by the dithering process, such that it does not yield
detectable sources.
We also note that no correction has been applied for non-linearity:
the individual exposures are below the threshold for non-linearity
effects, and only few, very bright sources are detected in individual
exposres, the others being within the background noise. We therefore
expects that some residual non-linearity may at most affect only the
brightest sources that are not the scientific target of the survey.

\subsubsection{Background subtraction}
After pre-reduction, the image backgrounds are still far from flat. 
Structures appear both at small and at large scales due to a variety 
of causes, such as pupil ghosts, dust, and sky-background variation 
during the observation (which is particularly strong 
in the near-infrared, especially in $K$).  We have
developed specialised algorithms to carefully remove these
structures. Since they are assumed to be additive features, the basic
operation is to create and then subtract maps of the background from each
image. The first map is a sigma-clipped median-stacked image of the
observations included in a temporal window -- typically of about 10
minutes -- around the processed image.  $Y$-band images show 
a substantial improvement after just the subtraction of this first 
sky background map, while a further large-scale polynomial fit to the residual features  has been further 
subtracted from the $K$-band images. 
At the end of this stage, in which each Observing
Block has been processed separately, images are flat and ready to be
processed to create the final mosaic.

\subsubsection{Astrometric solution}
In order to perform the coaddition, an accurate astrometric calibration
has to be performed. In fact images show geometrical distortions
arising from the positional errors of each pixel due to many causes,
such as optical distortions, atmospheric refraction, rotation of
chips, non-integer dithering pattern, etc.  The procedure of
astrometric calibration consists of two basic operations: the
correction of relative linear offsets between exposures and the
refined absolute global calibration.  For each exposure, a {\sc Sextractor}
\citep{Bertin1996} catalogue is created, and the relative linear calibration between
exposures is achieved by correcting for the offsets between source
coordinates, computed by the cross-matching with a catalogue chosen as
reference.  The absolute calibration is done by providing an absolute
reference catalogue and correcting for distortions through the cross-matching of source
coordinates and by storing the final corrected solution
into the header of the images.  We use as reference images 
the CANDELS WFC3/IR images \citep{Galametz2013, Guo2013} for both the 
UDS and GOODS-S fields. This approach was applied to each individual image
separately.

\begin{figure}
  \centering
  \includegraphics[width=8.5truecm]{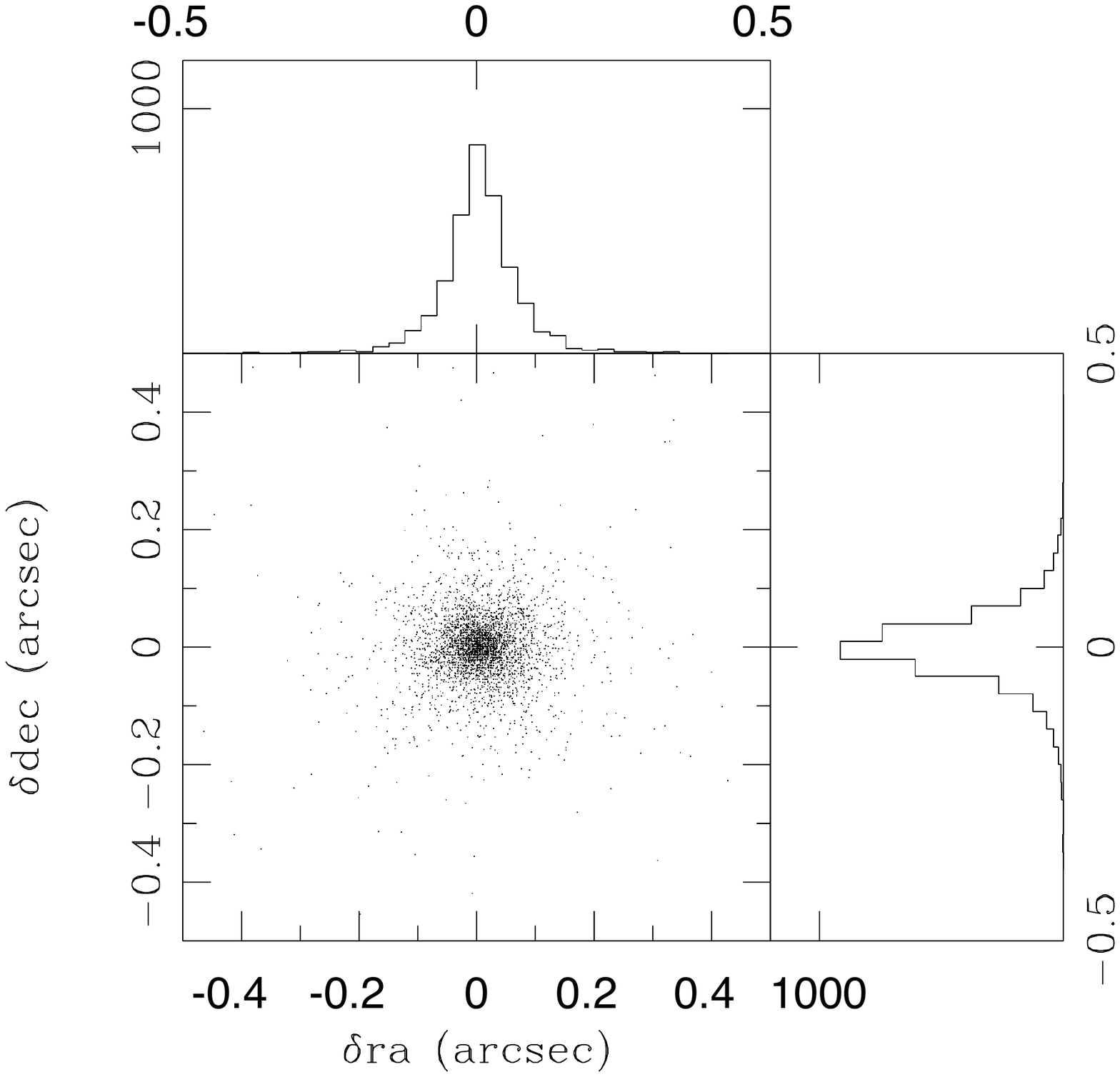}
   \caption{Differences in RA and DEC for all the objects detected in
     the overlapping region between the GD1 and GD2 pointings. The two
   insets on the right and top show the resulting distributions}
     \label{astrometry}
  \end{figure}

The final accuracy of the astrometric solution has been tested
using the regions with overlapping images. As an example, we show in
Figure~\ref{astrometry} the distribution of the differences in RA and DEC
for the objects that fall in the large overlapping area between the GD1
and GD2 pointings. The analysis is done on 3815  matched sources, and
the resulting r.m.s. are $\sigma_{RA}=0.067$'' and  $\sigma_{DEC}=0.074$''

\subsubsection{Estimate of absolute noise image}

Absolute noise maps for each exposure are created directly from the raw images. They are
based on the assumption that the noise is given by the Poisson
statistics of the counts detected in each pixel of the original frames. This contribution is
propagated to take the scaling applied to each pixel
during processing (including flat-fielding, normalization of exposure
times rescaling of zero-points etc)  into account. The resulting absolute noise map at pixel $X,Y$, $\sigma(X,Y)_{i}$ 
 can be obtained by the following formula:

\begin{equation}
\sigma(X,Y)_{i} =  \sqrt{\frac{RAW(X,Y)_{i}}{gain_{i}}} \times \frac{1.0}{\sqrt{FLAT(X,Y)_{i}}} \times \frac{1.0}{\sqrt{ndit}}
\end{equation}

\noindent where $i=1,2,3,4$ is an index that represents the number of the chip of HAWK-I, $RAW_(X,Y){i}$ the raw number counts at pixel $X,Y$,
$gain_{i}$ is the read-out gain, $FLAT(X,Y)_{i}$ is the value of the flat-field image used to calibrate the raw image, and $ndit$ is the value of $NDIT$.

\subsubsection{Coaddition}
Our pipeline uses SWarp (\cite{Bertin2002}) to resample all the processed images, 
implementing it into a procedure designed to properly propagate the absolute r.m.s. obtained as above.  
At first a global header is created from the input images that are resampled according to the geometry 
described in the resulting global header. In order to obtain a physical exposure and an r.m.s. 
map of the final mosaic, during the resampling the internal WEIGHT\_TYPE parameter has to be set to the MAP\_RMS modality, 
so that a noise map has to be given for each exposure.  During the resampling stage, 
images with bad astrometric information are rejected, while for each resampled image SWarp provides a weight map in output.  
The last step is to perform a weighted summation of all the resampled images, using as weight $w_i(X,Y)=1/\sigma_i(X,Y)^2$, 
where $\sigma$ for each pixel is given by Eq. 1.
The final r.m.s. map is obtained simply by: 
$ RMS(X,Y) = 1 / \sqrt{\sum _{i=1}^{n}{{w(X,Y)}^{i}}} $. These RMS images are released along with the science data (see below).
Since the current version of SWarp is not able to produce these r.m.s. images, these steps have been obtained with a specific pipeline.

\subsubsection{Photometric calibration}

At the end of each reduction we have adopted a careful procedure to
calibrate the photometry and estimate the zero point, independently
for each pointing.  For each pointing we have chosen at least one OB
qualified as photometric during the observations, and we have stacked
them in order to obtain a mosaic of the field with $\simeq 1$ hour of
exposure, in good photometric conditions and with consistent airmass.
We then retrieved from the ESO archive a set of standard stars observed at
the same airmass as the scientific images, and as close as possible in time 
to the observations.  We have reduced the
standards using the same calibration frames used for the scientific
images.  At the end of the reduction we have extracted a catalogue for
each standard calibration image and by comparing the magnitude of the
standard star, corrected for extinction, with the magnitude reported
in the literature, we have obtained a first estimate of the zero point $zp_1$
that we assume can be straight-forwardly applied to the stacked OB
described above.  Finally We have extracted and cross-matched two
catalogues, the first from the full complete mosaic, and the second from
the 1-hour mosaic, calibrated with $zp_1$. By comparing the
differences in magnitude between the sources in the two catalogues we
have achieved a final refined estimate of the zero-point.  To minimise
possible systematic offsets all these operations have been executed
using the {\sf MAG\_BEST} magnitude obtained by SExtractor on the
brightest objects only. We estimate that the typical uncertainty in
the derivation of the zeropoints is of $\pm 0.02$ {\it mag}, as obtained
from repeated estimate of the ZP on independent sets of OBs and
standards of the same pointing.

\section{Validation and tests on Photometry}
We have performed a number of tests, both on intermediate steps 
of data reduction as well as on the final images, comparing them with external data sets. 

We report here two classes of comparisons that may be of general
interest for the reader.
In all these cases we have obtained single-band photometric catalogues
using SExtractor and cross-matched the catalogues using the measured RA
and DEC. We then use the difference in observed total magnitudes for
the objects in common between the various catalogues. We note that the
stability of the photometric solution is in general quite good, to the
extent that its validation has been ultimately limited by the
uncertainties in the photometry. As our fields are relatively devoid
of stars, most of the objects that we have used for comparisons are
galaxies. It is well known that, when galaxies are observed with
different seeing, sampling and depth, the estimate of their total
magnitude is affected by systematics that depend on the size, profile
and surface brightness of individual objects. To minimise these
effects we have performed these tests on bright objects (typically
those detected at $S/N> 35$) and used Kron magnitudes, as measured by
SExtractor, that are relatively less sensitive to these effects.


First, we compared the catalogues obtained from fully-reduced stacks of
the different pointings in the overlapping regions. They
have revealed small differences (usually within the errors) of the
zeropoints that
have been averaged out in order to provide an internally-consistent data
set.



\begin{figure}
  \centering
  \includegraphics[width=7.5truecm]{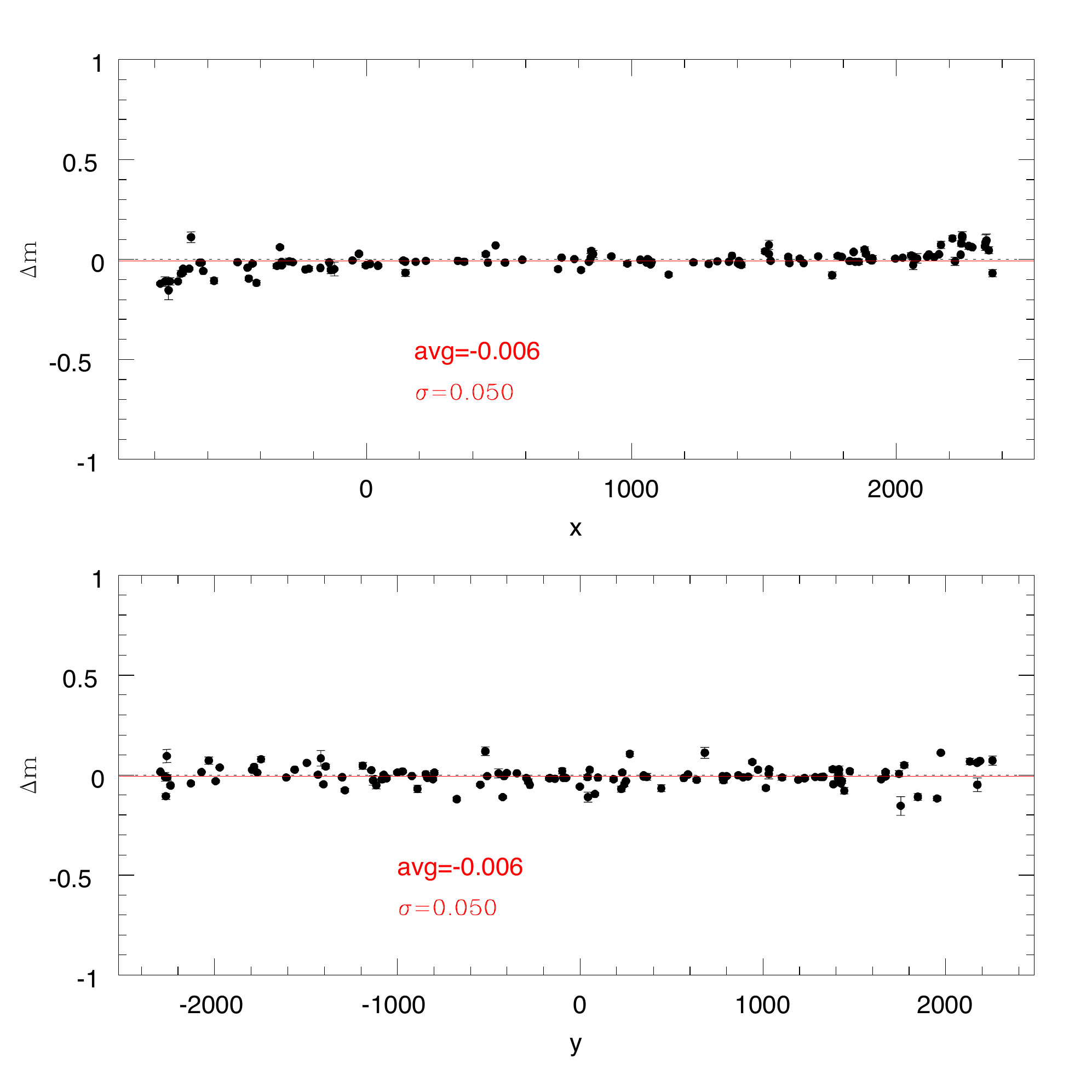}
 \caption{Difference in magnitude for objects detected in both
    the GD1 and GD2 fields, as a function of the X, Y position in
    pixels in the GD2 field. Since the field is rotated by $-$19.5
    degrees, using RA and DEC could hide real trends in the reduced
    data. This plot is before any renormalization of the zeropoint
    (see text for details).
       }
     \label{gd1_vs_gd2}
  \end{figure}

All the pointings of our survey have overlapping regions with (at least
one) nearby pointing. The typical size of these overlapping areas is
0.5''$\times$7' in UDS, and larger in GOODS-S (see Figure \ref{GOODSlayout}). We have used the
observed systematic offsets between the magnitudes in the various
pointings (as shown in Figure \ref{gd1_vs_gd2}) as a measure of the
systematic differences in the derived zeropoints. As shown in Figure
\ref{gd1_vs_gd2}, these offsets are always small (of the order of 0.01
- 0.02 {\it mag}), consistent with the uncertainty of the flux
calibration. For the UDS, we have tied the photometry to the zeropoint
of UDS2, i.e. we have renormalised the original zeropoints
of UDS1 and UDS3 in order to make the photometry of the overlapping
areas fully consistent. In GOODS-S we have similarly tied the
photometry to that derived from GD1.

We have also compared our final images to the previously available
images obtained by wider-field imagers. The goal of this exercise is
to further check against large-scale trends that may be left in our
data, that cannot be identified on the overlapping areas. 

\begin{figure}
 \centering
 \includegraphics[width=7.5truecm]{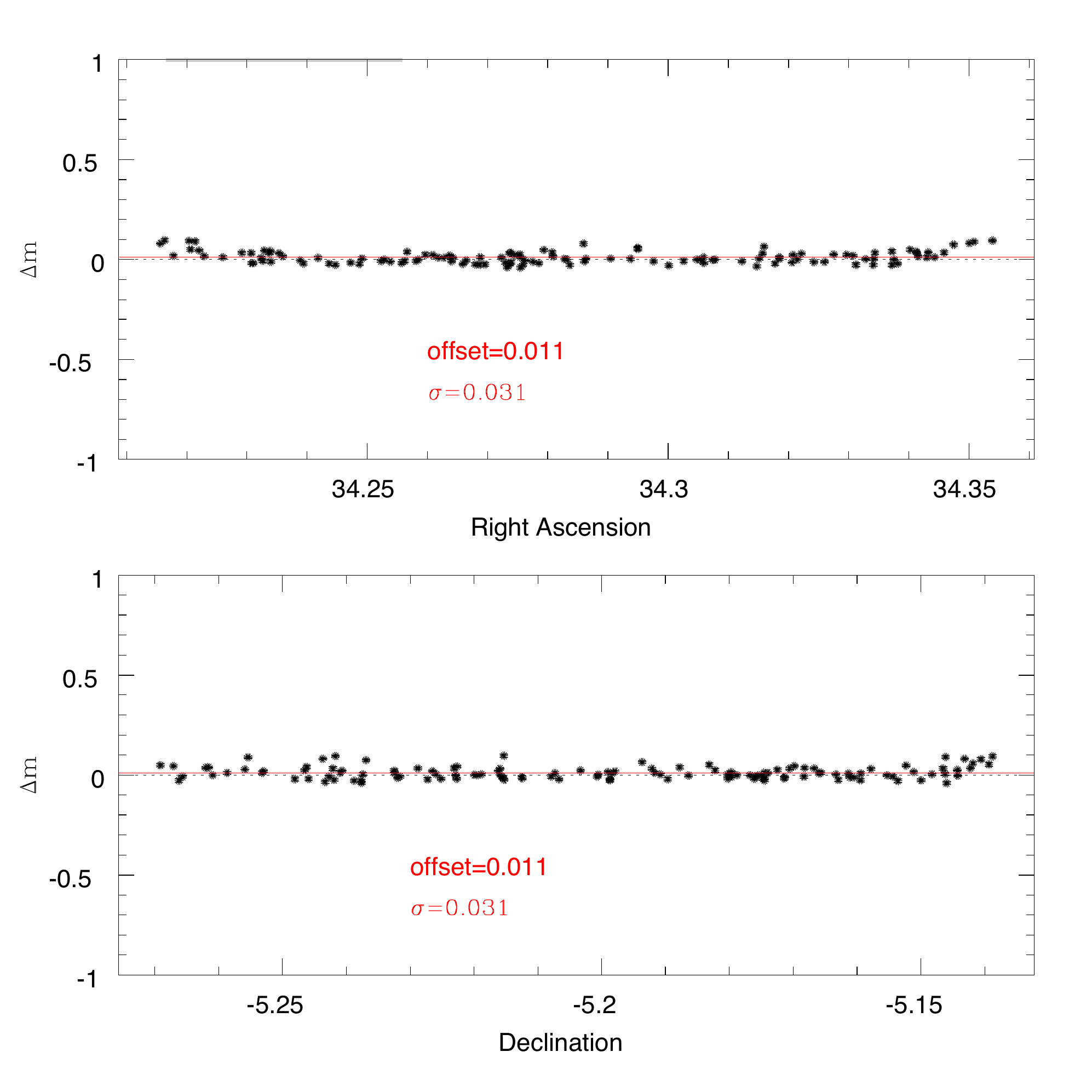}
\caption{Difference in magnitude for objects detected in the UDS2
  pointing and in the UKIDSS DR8 release, as a function of RA and DEC
   (see text for details).
      }
    \label{uds2_vs_ukidss}
 \end{figure}

\begin{figure}
 \centering
 \includegraphics[width=7.5truecm]{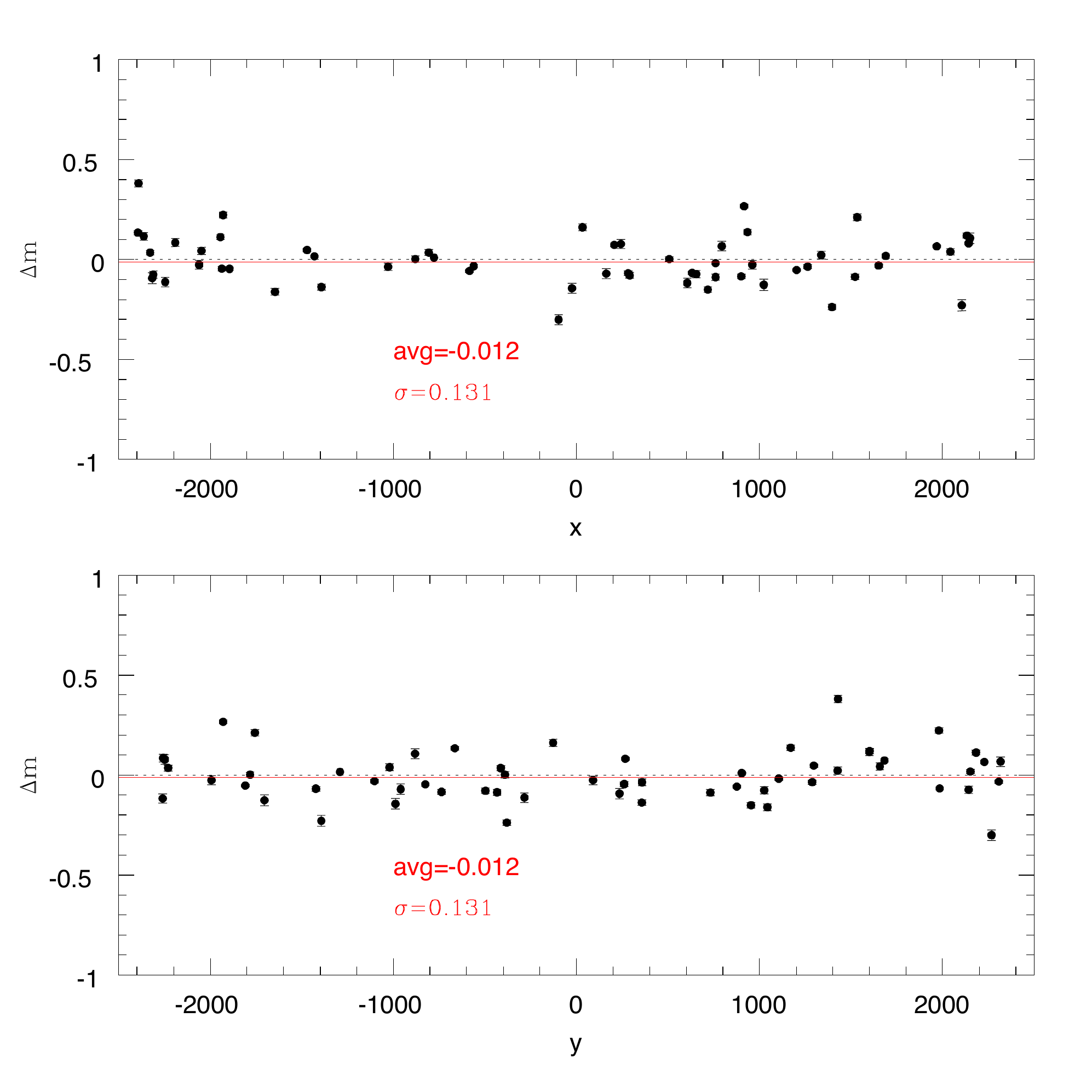}
\caption{As in Figure\ref{uds2_vs_ukidss}, for the GD2 and ECDFS field. As in Figure\ref{gd1_vs_gd2} we plot the difference  as a function of the X, Y position in
    pixels in the GD2 field. Since the field is rotated by $-$19.5
    degrees, using RA and DEC could hide real trends in the reduced
    data   (see text for further details).
 }
    \label{gd2_vs_ecds}
 \end{figure}

 For the UDS we have used the DR8 release of the UKIDSS Deep Survey,
 that has imaged a 0.77\,deg$^2$ field that includes the CANDELS
 field. For the GOODS-S we use the output of the Extended {\it Chandra} Deep Field (ECDFS), that was
 obtained with the SOFI instrument on NTT. Results are shown in
 Figures \ref{uds2_vs_ukidss} and \ref{gd2_vs_ecds}. Again, they
 show no major systematic trends in the data.

\section{Data release}
Both images and catalogues derived from the HUGS data are made publicly available. They can be retrieved from the ASTRODEEP 
website\footnote{http://www.astrodeep.eu/HUGS} as well as from the ESO,  {\it HST}, and CDS archives.

\subsection{Images}
 \begin{figure*}
  \centering
  \includegraphics[width=17truecm]{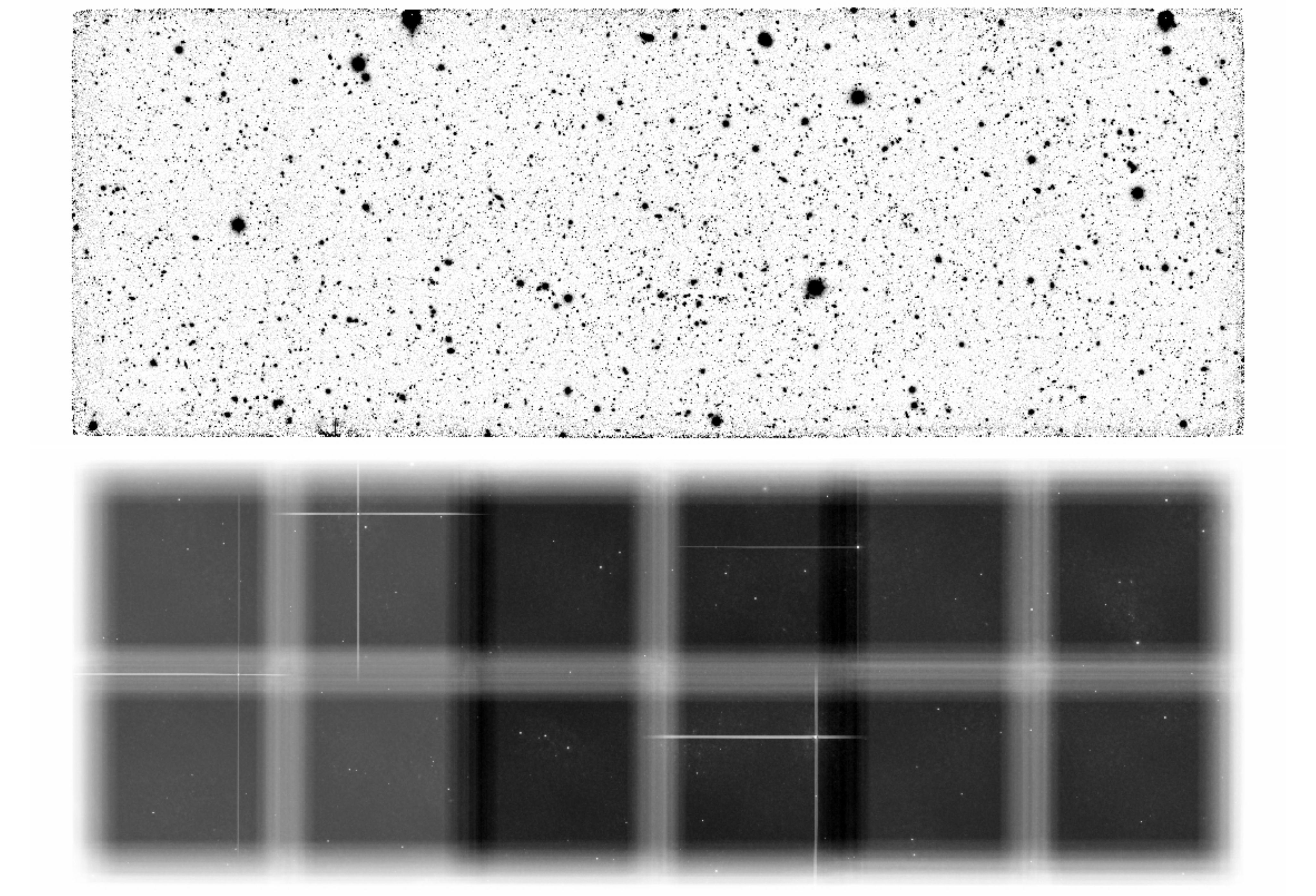}
  \caption{{\it Top: } Final image of the UDS field in the $Y$
    band. {\it Bottom: } Weight image, computed as described in
    the text. The bottom scale represents the r.m.s., normalised to
    its peak value. Darker regions have lower r.m.s., or equivalently
    higher relative weight, and hence correspond to deeper regions of
    the images. The left-most pointing (UDS3) is slightly shallower
    than the other two pointings, despite the fact they have the same
    exposure time, because the average background observed during the
    observations turned out to be larger than for the other two
    pointings.}
     \label{UDS_FINAL}
  \end{figure*}

  \begin{figure*}
  \centering
  \includegraphics[width=8truecm]{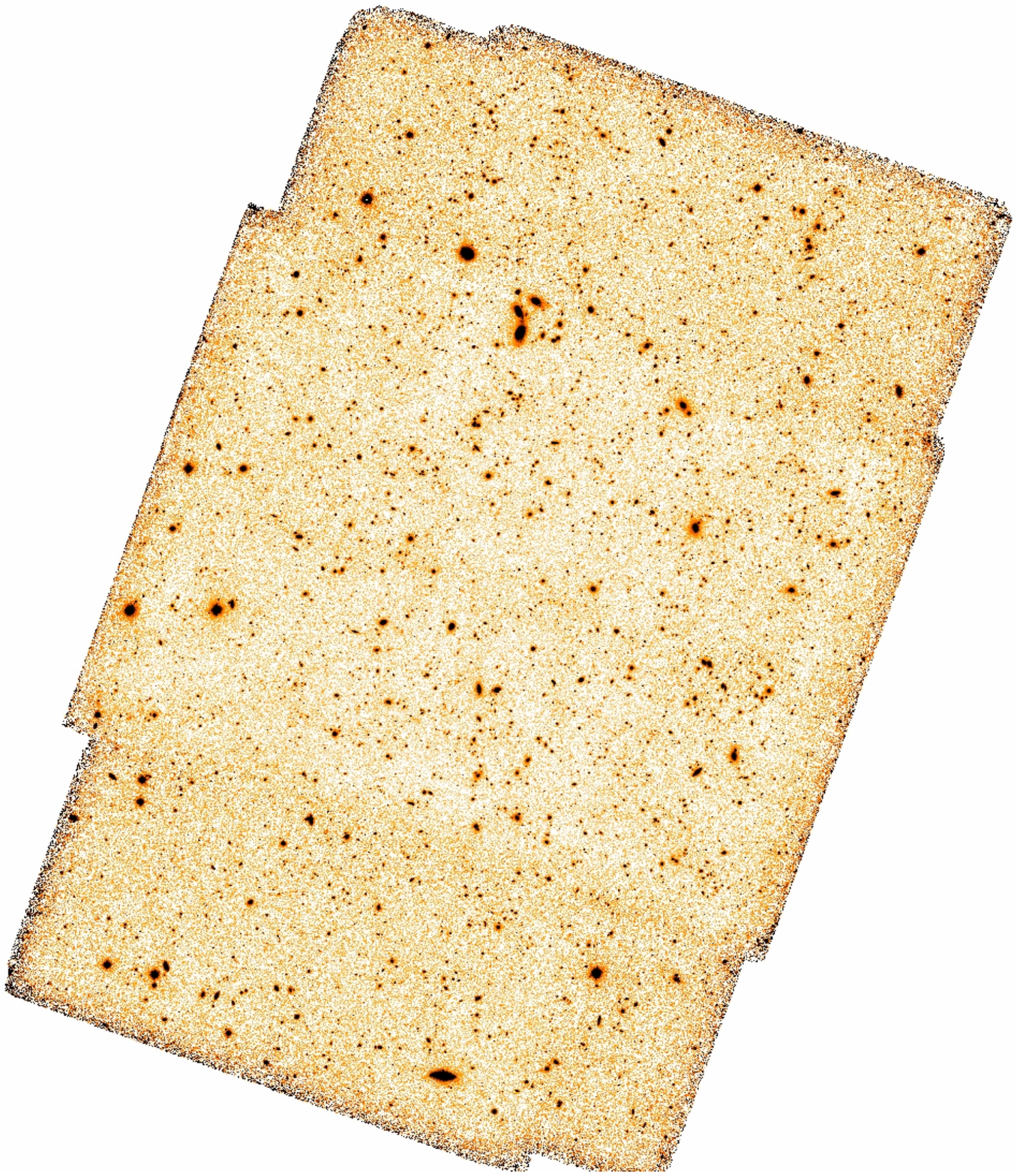}
  \includegraphics[width=8truecm]{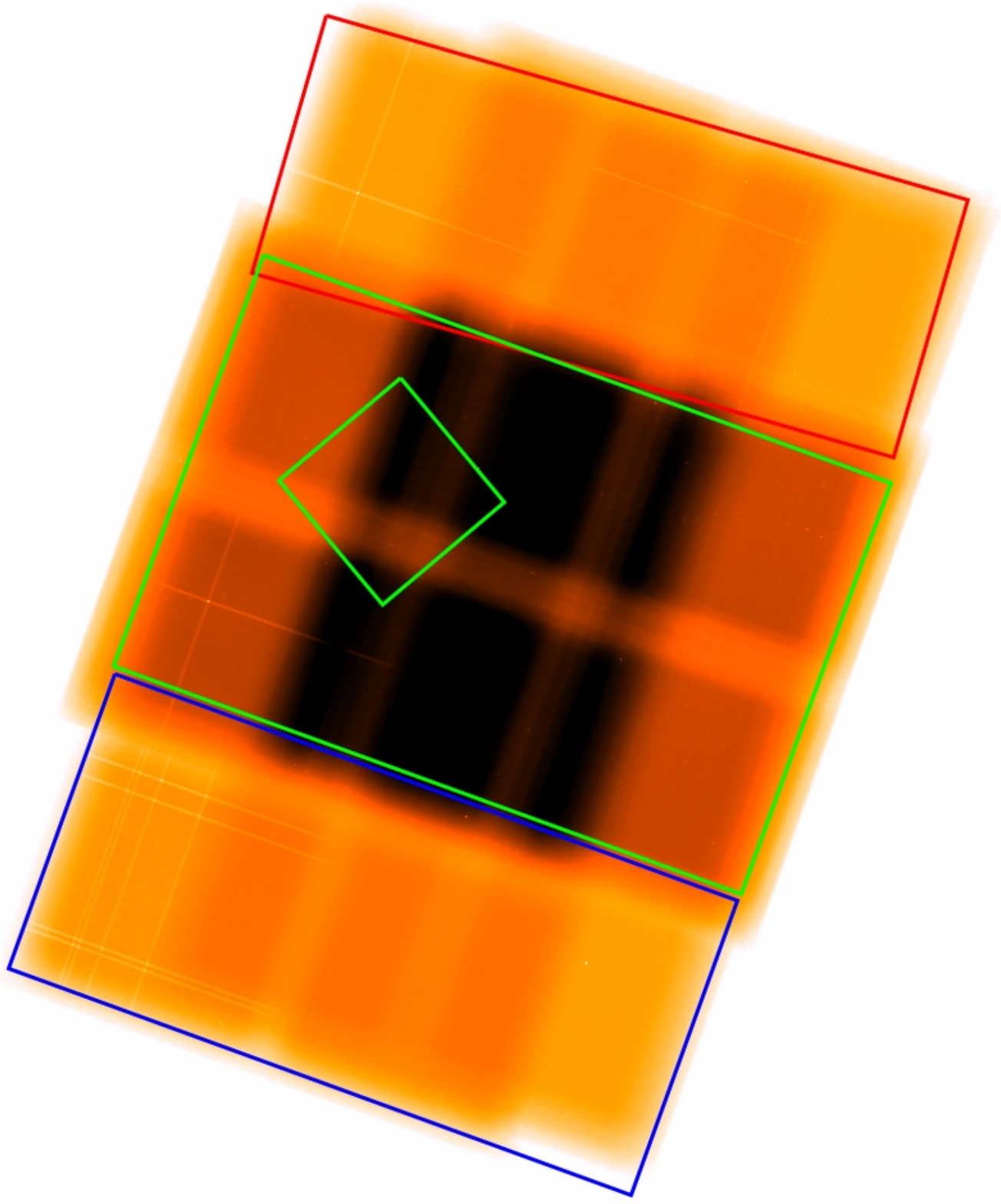}

  \caption{{\it Left: } Final image on the GOODS-South field, in the $K$
    band. {\it Right: } Weight image, computed as described in
    the text. Darker regions have higher weight - hence correspond
    to deeper regions of the images. }
     \label{GOODS_FINAL}
  \end{figure*}

  \begin{figure}
 \centering
 \includegraphics[width=8.5truecm]{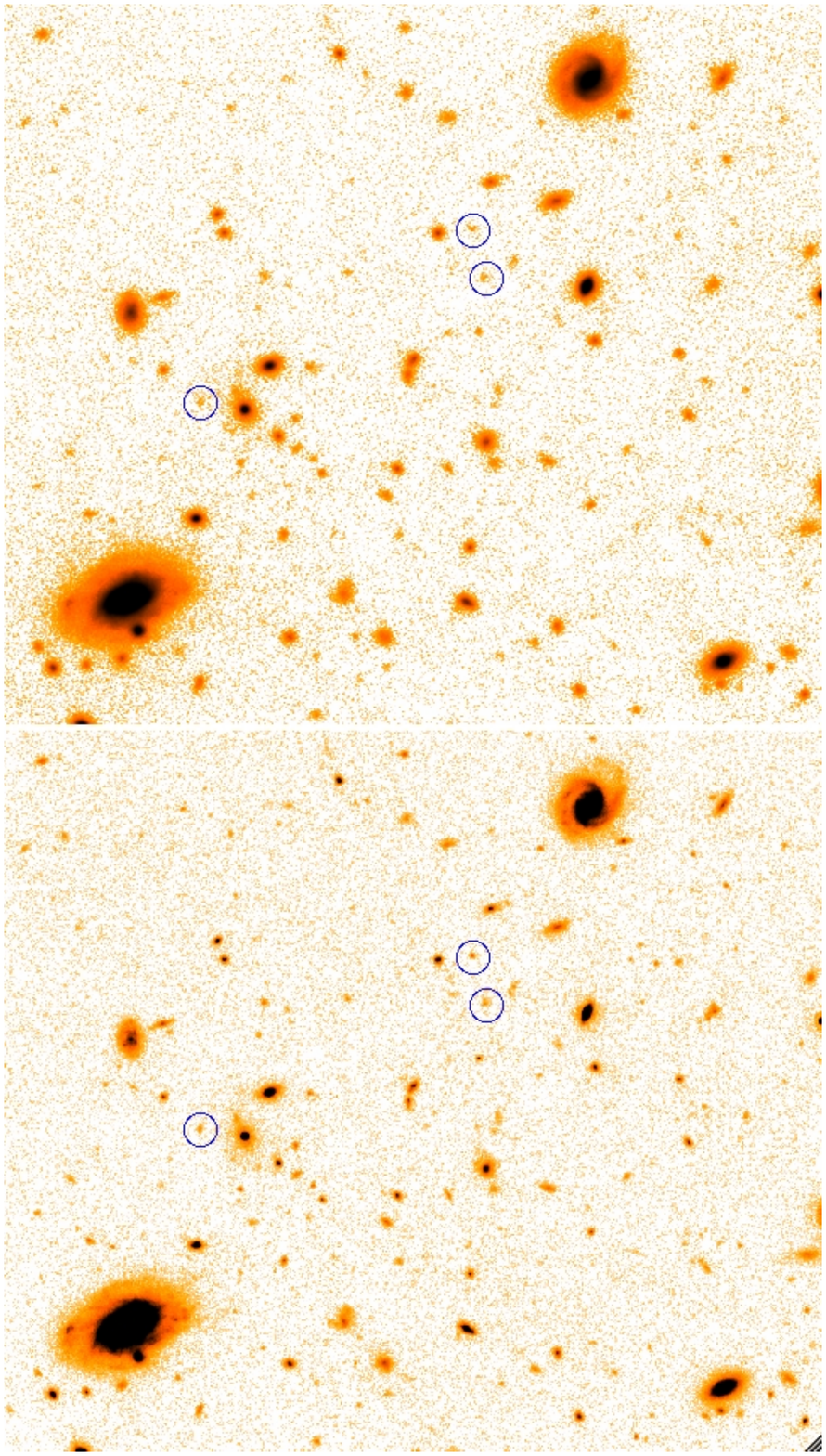}
 \caption{Centre of the GOODS-S field as observed with Hawk-I in
   the $K$-band (upper) and with WFC3/IR in the $H$-band (bottom). The displayed
   area is 1\,arcmin wide, and is extracted from the region where the
   Hawk-I data have the maximum depth. The $H$-band image is from the
   CANDELS Deep area.  Both images have a dynamic range extending
   from 0.5$\sigma$ to 100$\sigma$ per pixel, on a logarithmic
   scale. Objects encircled have $H_{160} \simeq 26$ and a colour
   $H-K\simeq 0.5$, typical of faint galaxies.}
    \label{hawki_wfc3}
 \end{figure}

  \begin{figure}
 \centering
 \includegraphics[width=8.5truecm]{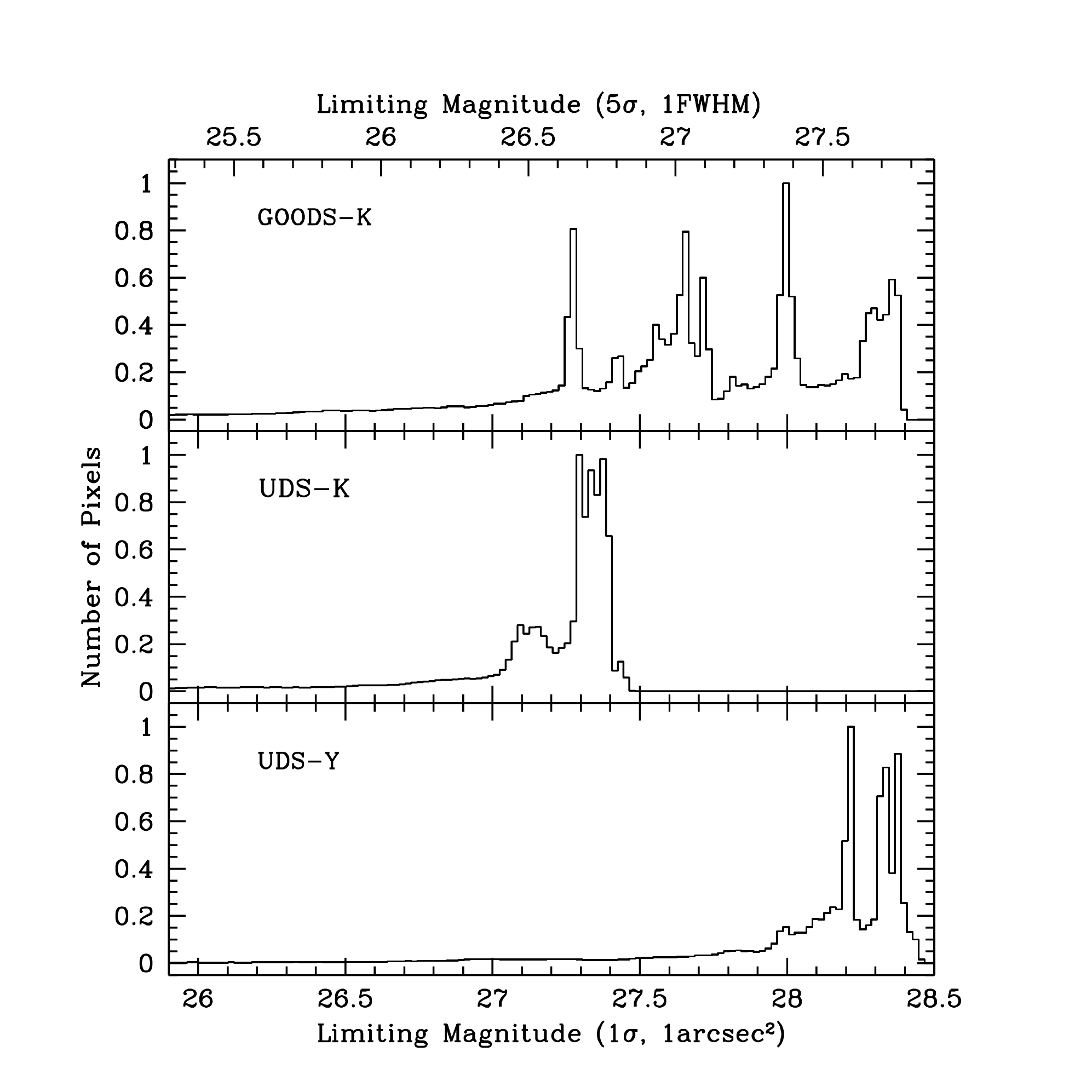}
  \caption{Distributions of limiting magnitude in the 3 HUGS images, as
    described in the legend. The lower x-axis shows the limiting magnitude  computed at
    1\,$\sigma$ in an area of 1\,arcsec$^2$, the upper at 5\,$\sigma$ in an
    area of 0.4\,arcsec$^2$, comparable to the average FWHM. Note that
    the latter is not the total magnitude of objects detected at
    5\,$\sigma$ in an area of 0.4\,arcsec$^2$, since no aperture
    correction has been applied.}
    \label{maglim}
 \end{figure}

We have finally obtained a set of mosaics of each data set, that we
make publicly available. 
For each data set we release:
\begin{itemize} 
\item  The coadded image for each pointing (UDS1,2,3,
  GOODS-D1, D2, W1, W2, W3 and W4) -- these are all calibrated and
  rescaled to a standard zeropoint of 27.5 for the $K$-band images, and 27.0
  for the $Y$-band;
\item The relevant absolute r.m.s. images, with the same flux scale;
\item A global mosaic of the two fields in each band, with the
  relevant absolute r.m.s., after homogenizing all images to the same
  PSF. This has been done by degrading the highest quality images to
  the one with the poorest seeing. Although the seeing is fairly constant
  and of excellent quality, this procedure inevitably degrades some 
  of the information contained in the data. The correlation of the
  background pixels is also visibly different across each image, because of
  the different degree of filtering applied to each original pointing.
\item  A global mosaic of the two fields in each band, with the
  relevant absolute r.m.s., without any correction for the different
  PSFs. These mosaics have varying PSF across the fields (in a smooth way 
across the overlapping region) but do not show a  varying degree of
correlation in the background. These are probably most useful for 
illustrative purposes, and have been used in the subsequent illustrations.
\end{itemize} 

We remark that the most accurate scientific analysis is, for most
application, obtained by separately using the individual pointings and
then combining the output in an appropriate way, as we did for the
derivation of the  multi--colour catalogues in the UDS (Galametz et al 2013, and see
below).

The final images for the two fields are shown in Figure~\ref{UDS_FINAL}
and Figure\ref{GOODS_FINAL} The images shown are those obtained
by combining the various pointings into a single mosaic, without 
performing any PSF matching prior to
coaddition. 

Figures \ref{UDS_FINAL} and \ref{GOODS_FINAL} show also the
r.m.s. images of the final mosaics. The relics of cosmetic defects are clearly visible in
the r.m.s. images, as are less exposed regions where some of
the data have been removed to eliminate defects or trails.

The comparison between the depth of the HUGS images and the 
CANDELS $H$-band images is performed more thoroughly in the next section 
using the multi--wavelength catalogues. In the meantime, a visual comparison is 
offered in Figure\ref{hawki_wfc3}, where we compare the deepest region from 
HUGS and CANDELS. It can be immediately appreciated that the depth and 
quality of the HUGS images is comparable in all aspects to the WFC3 data.

Because of the inhomogeneous depth of the final mosaics, the limiting
magnitude is not constant across the fields, in particular for the
GOODS-S imaging. This is shown in Figure\ref{maglim}, where we plot the
distribution of the magnitude limit in the two fields. This has been
computed converting the calibrated r.m.s. contained in each pixel into
a 1\,$\sigma$ magnitude limit in 1\,arcsec$^2$. The two peaks in the UDS
field come from the slightly shallow exposure obtained in the UDS3
pointing, where the sky background effectively observed in the data was
higher than the average in the other two. In GOODS-S, the 4 broad
peaks in the distribution of the magnitude limits come from the
complex geometry of the exposure, as shown in Figure\ref{GOODS_FINAL}.


\subsection{Catalogues}  
Catalogues from the HUGS images can be extracted in two different
ways, either as ``single band'', i.e. using the HUGS images as
detection image, or adding them to the full multi--wavelength suite of
data in CANDELS. 

\subsubsection{Single band catalogues} 

Single-band catalogues are, in principle, straightforward to obtain. 
We have used the SExtractor code to obtain
single-band detected catalogues that are distributed along with the
images. However, a compromise must be achieved in this case between
two conflicting requirements. First, to make full use of the largest
possible depth, the detection should be made on the global mosaic,
especially in GOODS where the overlaps between the various pointings
are significant. However, since the detection process must be tuned to the
PSF of the images, and given the different seeing within the HUGS images,
to obtain a fully-consistent catalogue we have used as input images the
seeing-homogenised images presented above. This somewhat limits the
possibility of detecting the faintest galaxies in the pointings with
the best seeing, although the seeing variation among the various
images is not dramatic (see Tables \ref{UDStable} and \ref{GOODStable}). 
We deemed this procedure as the most appropriate to obtain
single-band catalogues, that are made publicly available for future
use.  For the same reason, the number counts presented in the next
section have been obtained with a slightly different procedure that we
describe in the next section.
Summarizing, the two public catalogues that we derive have been obtained
using SExtractor on the seeing--averaged mosaics of both the UDS and
GOODS-S HUGS imaging. We have applied a smoothing before detection (of
the same size of the PSF), and used a
minimal detection area of 9 pixels, requiring $S/N>3$ in such an area.

\subsubsection{Multi--wavelength catalogues} 

We have included the photometry extracted from the HUGS 
images (both in $K$ and in $Y$) in the UDS and GOODS multiwavelength catalogues
described in \citet{Galametz2013} and \citet{Guo2013} respectively. In both cases
we have detected the objects in the WFC3 $H$-band image from CANDELS,
and performed PSF-matched photometry on the HUGS images. 
This has been accomplished by using the TFIT package \citep{Laidler2007} to properly 
take the morphology of each object into account  during the deblending process. We refer
the reader to those two papers for more details. The Guo et al. catalogue was compiled 
using only the first epoch of Hawk-I images in GOODS-S. We have therefore re-extracted
the $K$-band photometry using the final images described here, for all the objects 
detected in the $H$-band. This catalogue is released here.

\begin{figure}
 \centering
 \includegraphics[width=8.5truecm]{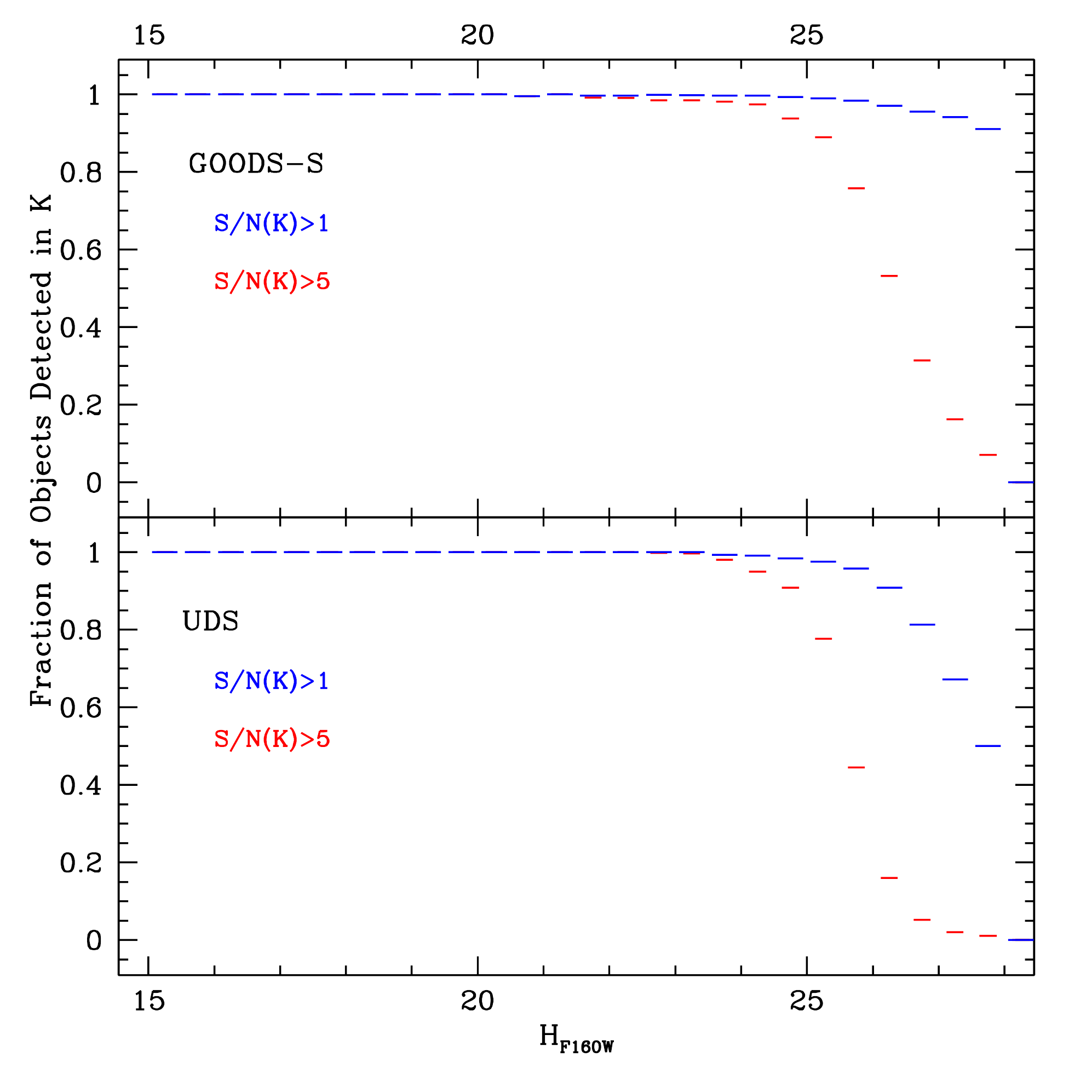}
  \caption{Fractions of objects in the CANDELS catalogues that
  have a detected flux in the HUGS data, as a function of the $H$-band magnitude. 
Here the $H$-band is measured on the CANDELS F160W images, and
the corresponding $K$-band flux is measured with the TFIT code on the
final HUGS images. Results are shown for two different signal--to--noise ratios in
the $K$-band, and for the two HUGS fields (as shown in the legends). Errors are computed assuming simple Poisson statistics}
    \label{detfrac}
 \end{figure}

We note that, to deal with the different PSFs of the various images,
we have independently processed each of the final individual pointings
described above, and thereafter weight-averaged the photometry of
objects detected on multiple images to obtain the final
photometry. Clearly, following this procedure, the ultimate depth of
the catalogue is driven by the WFC3 $H$-band image.  It may be of some
interest to show how effective the HUGS images are in providing us
with useful information on the CANDELS-detected objects, since that is
one of the main aims of the HUGS survey.  This is shown in Figure
\ref{detfrac}, where we plot the fraction of objects that are detected
at $5\sigma$ or $1\sigma$ in the $K$-band as a function of the
$H$-band magnitude.  In the case of GOODS-S, it is seen that as many
as 90\% of the $H$-detected galaxies have some flux measured at
$S/N>1$ in the $K$-band, down to the faintest limits of the $H$-band
catalogue, and that nearly  60\% of the $H \simeq 26$ galaxies (and
15\% of the $H \simeq 27$ galaxies) have a solid $K$-band detection
with $S/N>5$. We note that these statistics are measured on the full
GOODS-S HUGS area, which is highly inhomogeneous in depth both in the
$H$ and $K$ bands.  This result confirms that our original goal has
been achieved, and in particular that our pointing strategy has been
quite efficient in covering the inhomogeneous GOODS-S field at the
required depth.

\section{Number Counts} 

 \begin{figure}
 \centering
 \includegraphics[width=8.5truecm]{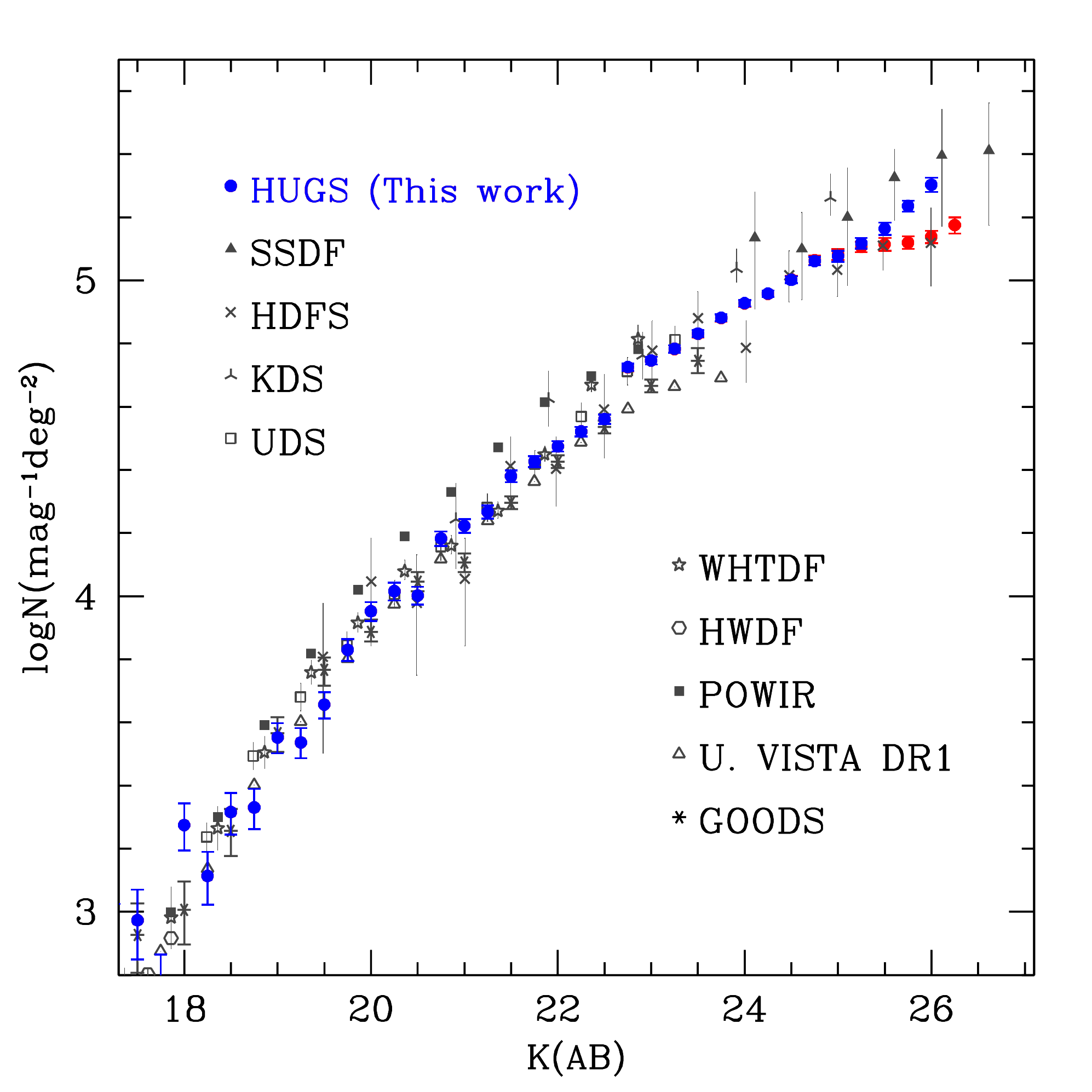}
  \caption{Number counts in the $K$-band, from HUGS and from the recent
    literature (see text for full references for the previous surveys). 
    In blue are the HUGS counts corrected for incompleteness
    assuming a faint galaxy size distribution spanning the range $r_{hl} =0.1 -0.3$\,arcsec. In red we show
    the counts that are derived assuming point-like sources. Recent
    results from the literature are also shown: SSDF: \cite{Minowa2005}, HDFS:
\cite{Labbe2003}, KDS: \cite{Moustakas1997}, UDS: \cite{Cirasuolo2010}, WHTDF: \cite{Cristobal-Hornillos2003},
HWDF: \cite{Huang1997}, POWIR: \cite{Conselice2008},U.VISTA DR1: \citep{McCracken2012}, GOODS: \cite{Grazian2006}.}
    \label{counts}
 \end{figure}

 Finally, we have derived the number counts in the $K$-band, combining both 
 UDS and GOODS-S images. At variance with the procedure described
 above, we decided not to use the seeing-homogeneised mosaics since
 their use would limit the final depth of the catalogues (due to the
 compromised seeing). Within the UDS field we therefore extracted 
independent catalogues from each of the three UDS Hawk-I pointings (which have notably different seeing),
 optimizing each catalogue to the relevant PSF. 
 In the case of GOODS-S, we built a specific mosaic using the two deep
 pointings D1 and D2 as well as W2 and W3. These four pointings have
 remarkably similar seeing, thus enabling us to build a mosaic without any
 additional seeing-homogenization, which allows us to exploit the deepest images
 obtained in HUGS without degrading their quality.

Four catalogues (UDS1, UDS2, UDS3, GOODS-D1+D2+W2+W3) have thus been obtained 
using SExtractor, as described above.
As already mentioned we use the Kron magnitudes (termed MAG\_AUTO in SExtractor), and  
the systematic errors associated with their use are corrected with the simulations 
that we describe below. Number counts have been derived independently from each image using the
procedure described below, and a weighted average of these has then been used 
to obtain the final  number counts. The small overlap between
the UDS fields means that these three catalogues are not entirely independent, 
but  we have ignored this effect since the objects detected in more than one image 
represent only $\simeq 1$\% of the total sample. 
After trimming the outer regions of the images, the final area over
which we compute the number counts is 340.58 square arcmin, i.e. about 1/10th of a square degree.
We note however that the area over which the number counts are measured is in practice 
a function of magnitude because of the inhomogeneity of our exposure maps, As a result,
the deepest number counts (i.e. those determined at $K>25$) are in practice determined 
from a sub-area of 50.17 square arcmin within the GOODS-S field. 

Central to a robust determination of the number counts is a thorough and reliable estimate of 
the incompleteness and other systematics/biases, which must be obtained through the use of simulations. 
As is now standard procedure, we have performed these simulations by inserting fake objects into the 
real images with a range of magnitudes (from $K=18$ to $K=27.5$) and sizes (we used exponential profiles 
with observed half-light radius chosen randomly from within the range $0 - 0.4$\,arcsec). 
These objects have also been convolved with the observed PSF before being placed at random positions 
in the field, including areas where known objects are already detected. Object detection 
is therefore achieved in the very same way as for the real objects in the original images,
 and the output magnitudes measured for the fake objects (when detected) are measured, along 
with the detected fraction. Only 200 objects have been placed in each run, in order to minimise 
excessive and unrealistic crowding effects. Simulations have been repeated 
until $10^6$ objects have been simulated in each field.

To use these simulations, we have followed exactly the same approach used in the analysis 
of the deepest $K$-band image obtained so far, in the AO--assisted Subaru Super Deep Field 
(SSDF, \cite{Minowa2005}; the technique is similar to the one adopted
also by previous analysis, e.g. \cite{Smail1995}).  This method takes into account three sources of systematic error. 
The first is the incompleteness, i.e. the fraction of objects lost as a function of their real input magnitude. 
The second is the systematic bias that can arise in the estimate of total magnitude due to surface brightness 
sensitivity. Specifically,  
at low signal--to--noise ratins, the Kron magnitudes that we use (MAG\_AUTO in SExtractor) progressively underestimate 
the real magnitudes of the detected galaxies; our simulations indicate that at faint fluxes this 
effect can easily be $0.1-0.15$\,{\it mag}, and neglecting it would bias the estimate of the slope of the number counts.
Finally, there is the effect of flux-boosting; fainter sources can be   
promoted to brighter magnitudes (hence contaminating the number counts)
when they happen to fall on a positive noise fluctuation.
Technically, this is done by first extracting from the simulations the transfer 
matrix $T_{oi}$ that gives the fraction of galaxies with input magnitude $m_i$ that are 
detected with output magnitude $m_o$. We then build the probability matrix $P_{io}$ 
that gives the probability that an artificial object detected with $m_i$ actually has a 
magnitude $m_o$. The elements of $P_{io}$ are computed as $P_{io} = T_{oi}n_i / \sum_k{T_{ok}n_k}$, 
where $n_i$ is the number of artificial objects with magnitude $m_i$. In building 
this matrix the relative fraction of input galaxies must scale with a realistic slope, in order to 
appropriately weight the number of galaxies close to or fainter than the magnitude limit. 
To ensure this is done properly, objects are simulated up to $K\simeq 27.5$, 
much deeper than the formal 5-$\sigma$ detection limit of even the deepest regions 
of our Hawk-I imaging. This is done assuming the true intrinsic number counts 
follow a power-law whose slope is varied until a consistent result is achieved 
with the recovered number counts.  The final number counts are finally computed 
as $n^{cor}_i = \sum_o{P_{io}n_o^{gal}}$, where $n_o^{gal}$ are the raw (observed) number counts.  We defer to
\cite{Minowa2005} for more details.

The estimate of the size and magnitude dependence of the correction depends on a 
critical assumption, namely the distribution of galaxy sizes. At the exquisite resolution 
of the HUGS images the difference between compact and point-like sources is measurable, 
as in the \cite{Minowa2005} data.  To get an estimate of the real size distribution of 
the galaxies at faint $K$-band magnitudes, we have looked at the half-light radius 
($r_{hl} $) of galaxies as measured by SExtractor in the $H$-band WFC3 images in CANDELS-GOODS. 
While the $r_{hl} $ of stars is 0.15\,arcsec (corresponding to 
the angular resolution limit of $H$-band {\it HST} imaging), 
we find that galaxies at $K\simeq 26$ (the typical magnitude where incompleteness is effective 
in the deepest GOODS data, see below) have $r_{hl} $ typically in the range $0.15 - 0.3$\,arcsec. 
There is also a clear trend with magnitude, with brighter galaxies being even larger 
than $0.3$\,arcsec, while the typical $r_{hl}$ for galaxies at $K\simeq 27$ appears to be much closer 
to 0.15\,arcsec (the value indicative of unresolved objects).  We therefore computed 
the correction for incompleteness in two cases: i) assuming point--like sources 
(as in \cite{Minowa2005}) and ii) assuming a distribution in size between 0.1 and 0.3\,arcsec. 
We performed this exercise independently in the various images for both fields and then 
averaged the resulting number counts, weighting them by the area of the appropriate parent image.  
Finally, to minimise the sensitivity of our derived number counts to these corrections, we have used the counts from the various 
images only when the incompleteness is negligible ($<5$\%), the only exception being the deepest areas in GOODS-S where 
we have of course used our best estimate of the appropriate corrections to push our 
derived number counts to the faintest limits.

The derived number counts are given in Table\ref{TableCounts}. 
They are also shown in Figure~\ref{counts}, for both assumptions about galaxy size,
compared with a number of recent results from the literature (SSDF: \cite{Minowa2005}, HDFS:
\cite{Labbe2003}, KDS: \cite{Moustakas1997}, UDS: \cite{Cirasuolo2010}, WHTDF: \cite{Cristobal-Hornillos2003},
HWDF: \cite{Huang1997}, POWIR: \cite{Conselice2008},U.VISTA DR1: \citep{McCracken2012}, GOODS: \cite{Grazian2006}).
Uncertainties have been estimated assuming simple Poisson errors.  

As expected, the number counts agree very well with previous results from
the literature. It is immediately appreciated that the HUGS number
counts exceed in depth and statistical accuracy all previous estimates
at the faint magnitudes, the only exception being the very faintest bin 
plotted from the SSDF. The latter used AO-assisted observations to achieve a very small
PSF (0.18\,arcsec), making these observations more sensitive than HUGS to
the detection of very faint point-like sources.  The faint counts from SSDF 
are, however, very uncertain, although we note that they are consistent 
with the HUGS counts derived assuming extended galaxy morphology.

At the faint limit, it is immediately clear how dramatic the impact of
the assumptions on galaxy size is. Assuming point-like sources we
confirm the flattening of the number counts at $K \simeq 24-26$ reported by 
\cite{Minowa2005}, with a slope $d\log N / dm$ of about 0.15, in our case with a much larger statistical
accuracy (due to the 50$\times$ larger field of view of our images).
However, assuming instead a typical galaxy size in the range $0.1-0.3$\,arcsec, we
find that the slope of the number counts remains essentially unchanged
up to $K\simeq 26$, with a slope of about 0.18.

This difference in slope is potentially very important for establishing the contribution of detected galaxies to the extragalactic background light (EBL).  As noted by \cite{Minowa2005}, a relatively shallow slope at faint limits (as found assuming point--like sources) would lead to a major discrepancy with the observed EBL, leaving space for additional sources, for instance primeval galaxies. Conversely, if we assume that the steep slope that we find for resolved galaxies holds to even fainter limit, the consequence is that most of the diffuse EBL observed from space can be ascribed to ordinary galaxies, without the need to invoke for more exotic populations.

A robust determination of this effect, however, requires more refined simulation 
and analysis of the galaxy size distribution as a function of magnitude, possibly including 
even deeper data such as is provided in the $H$-band by HUDF12. 
However, such an analysis is beyond the scope of this current paper.

\begin{table} \caption{$K$-band galaxy number counts derived from the HUGS survey, 
corrected for incompleteness and other systematic effects as described in the text. 
Number densities are in units of galaxies per magnitude per square degree. 
The full table is available at the CDS and at http://www.oa-roma.inaf.it/HUGS}
\centering
\label{TableCounts}
\begin{tabular}{c c c c c   } 
\hline\hline

Magnitude bin & log (N)\tablefootmark{a} & log ($\sigma$(N))\tablefootmark{a} & log (N)\tablefootmark{b} & log ($\sigma$(N))\tablefootmark{b} \\
\hline
24.00 & 4.9274 & 3.2965 & 4.9283 & 3.2969  \\ 
24.25 & 4.9582 & 3.3285 & 4.9588 & 3.3288  \\ 
24.50 & 5.0033 & 3.4016 & 5.0021 & 3.4010  \\ 
24.75 & 5.0639 & 3.5820 & 5.0619 & 3.5810  \\ 
25.00 & 5.0827 & 3.6783 & 5.0774 & 3.6757  \\ 
25.25 & 5.1068 & 3.6904 & 5.1187 & 3.6964  \\ 
25.50 & 5.1147 & 3.7863 & 5.1648 & 3.8114  \\ 
25.75 & 5.1203 & 3.7891 & 5.2363 & 3.8471  \\ 
26.00 & 5.1391 & 3.7985 & 5.3038 & 4.0063  \\ 
26.25 & 5.1758 & 3.9423 & - & - \\ 
\hline
\end{tabular}

\tablefoot{
\tablefoottext{a}{assuming unresolved morphologies for galaxies; see text for details}
\tablefoottext{b}{assuming a distributions of galaxy half-light radius from 0.1\,arcsec to 0.3\,arcsec; see text for details} 
}
\end{table}

\section{Summary}
We have present in this paper the results of a new, ultra-deep,
near-infrared imaging survey executed with the Hawk-I imager at the
ESO VLT. This survey, named HUGS, provides deep, high-quality imaging in the $K$
and $Y$ bands over the portions of the UKIDSS UDS and GOODS-South
fields covered by the CANDELS {\it HST} WFC3/IR survey.  While the
bulk of the data presented here comes from a program specifically
designed to cover CANDELS, (the ESO Large Program 186.A-0898
P.I. Fontana) we have also included in our analysis other data coming
from previous observations on GOODS-S, which were acquired either
during the Science Verification Phase (60.A-9284)or in the framework
of another program designed to look for $z\simeq 7$ galaxies (The ESO
Large Program 181.A0717, P.I. Fontana,
\cite{Castellano2010a,Castellano2010b}). These data comprise nearly
all the available \Hw data on GOODS-S and have been homogeneously
reduced, discussed and made publicly available here.

This paper describes the survey strategy, 
the observational campaign and the data reduction process. 
For further details on the data reduction we refer the reader to the previous sections, 
which give full details. We simply mention here that we have followed standard 
recipes for these images, adopting a number of validation controls. 
First,  we have used two independent pipelines (one developed in Rome and one in Edinburgh) 
to reduce the first epoch of data, and cross-compared the results, finding excellent consistency 
and agreement. In addition, we used the large wealth of independent images acquired over each field 
to internally validate the data. At the end of these tests, we are confident that the 
observation uncertainties are under control, typically at the level of few percent. 

Similarly, full details of the observing strategy and resulting data are described in the text. 
We refer the reader in particular to the Tables and Figures for further information. 
We summarise here the fundamental aspects of our survey.

$\bullet$ HUGS has been designed to complement the CANDELS data in the two fields where 
crucial infrared data of the necessary depth cannot or have not been obtained with WFC3: 
deep $K$-band imaging in GOODS-S, and deep $Y$ and $K$-band imaging
in the UDS.  The depth has been tuned in order to match the depth of the WFC3 $J$ and $H$ imaging. 
For instance, the $K$-band limit is set about 0.5 {\it mag} shallower than $H_{160}$ senstivity limit, in order
to match the average $H-K$ colour of faint galaxies.\\
$\bullet$ Pointings have also been optimised to cover CANDELS. In the UDS, we covered 85\% of the CANDELS area with 
3 different, marginally-overlapping \Hw pointings. For the $K$-band in  GOODS-S we adopted a more complicated pattern, 
comprising 6 different pointings (two ``Deep'' and four ``Wide'') as described in Figure\ref{GOODSlayout}. 
This has allowed us to vary the exposure time over the field in order to match the varying depth of the CANDELS images. 
The $Y$-band in GOODS-S covers about 60\% of the eEastern CANDELS area.\\
$\bullet$ In the UDS, the exposure times of each pointing are $\simeq 13$\,hours in the $K$-band and $\simeq 8$\,hours in the $Y$-band. 
The seeing is $0.37-0.43$\,arcsec in the $K$-band and $0.45-0.5$\,arcsec in the $Y$ band. 
The corresponding limiting magnitudes are  $m_{lim}(K)\simeq 26$,  $m_{lim}(Y)\simeq 26.8$ (5$\sigma$ in one FWHM) 
or  $m_{lim}(K)\simeq 27.3$,  $m_{lim}(Y)\simeq 28.3$ (1$\sigma$ per arcsec$^2$).\\ 
$\bullet$ In GOODS-S, the total exposure time in the $K$-band (summed over six pointings) is $\simeq 107$\,hours. 
Because of the complex  geometry, this corresponds to an exposure of 60--80 hours in the central area 
(the sub-region covered by CANDELS ``Deep'' ) and 12--20 hours over the remainder 
(the CANDELS ``Wide area''). The final average seeing is remarkably good and constant, with 4 pointings with FWHM $\simeq 0.38$\,arcsec (notably including the two deepest) 
and 2 pointings with FWHM$\simeq 0.42$\,arcsec. On the final stacked images, the limiting 
magnitudes in the deepest area are   $m_{lim}(K)\simeq 27.$ (5$\sigma$ in one FWHM) or  $m_{lim}(K)\simeq 28.3$,  $m_{lim}(Y)\simeq 28.3$ (1$\sigma$ per arcsec$^2$). \\
$\bullet$ We have derived from the HUGS images two different types of catalogue - a $K$-band selected catalogue 
that we use to estimate number counts, and PSF-matched catalogues for all the $H$-selected galaxies detected in the 
CANDELS images. For the UDS, the corresponding catalogue published in Galametz et al. (2013) 
already uses the full HUGS data. However, the catalogue for GOODS-S obtained and made available here 
supercedes the catalogue already published in Guo et al. (2013, which used only a preliminary version of the HUGS data, 
obtained from shallower observations of the central area (pointings D1 and D2) only).\\
$\bullet$ Our key and crucial goal of effectively matching the CANDELS depth (for most useful analyses) has been 
achieved, as shown in Fig. \ref{detfrac}, where it can be seen that a large fraction of even the faintest $H$-band CANDELS objects 
are also clearly detected in our HUGS $K$-band imaging.

\smallskip

Finally we have presented the galaxy number counts in the $K$-band, as obtained after combining 
the two HUGS fields. We describe the simulations that we have adopted to correct for the 
incompleteness and flux boosting at the faintest limits, making different assumptions for the size 
distribution of faint galaxies.
We show that our number counts extend to magnitude limits fainter than any previous survey, 
other than the rather uncertain faint counts achieved with the 
AO-assisted images of the SSDF field. The latter have a FWHM of 0.18\,arcsec and exposure times comparable to our survey, and hence 
reach fainter limits for unresolved objects, but over such a small area ($\simeq 1$\,square arcmin) that they are statistically 
very uncertain. We show that the slope of the number counts over the faintest bins depends sensitively on the 
assumed distribution of galaxy sizes. It ranges from 0.18, if we assume that galaxies at $K>26$ are unresolved, 
to 0.22, if we assume that such galaxies have a typical distribution of half--light radii spanning the range
$0.1 - 0.3$\,arcsec, as suggested by a preliminary analysis of the deepest CANDELS images. 

We have made all of the final HUGS images and derived catalogues publicly available at the ASTRODEEP website \footnote{http://www.astrodeep.eu/HUGS} as well as from the ESO archive.

\begin{acknowledgements}
This work would not have been possible without the support and dedication
of the whole ESO staff. We thank in particular our support astronomer
Monika Petr-Gotzens. We are also grateful to the memory of Alan Moorwood, 
who was fundamental in motivating the development of the Hawk-I instrument, with which our survey 
has been undertaken. We also thank the referee, V. Manieri, for his accurate report.
AF and JSD acknowledge the contribution of the EC FP7 SPACE project ASTRODEEP (Ref.No: 312725).
JSD also acknowledges the support of the Royal Society via a Wolfson Research Merit Award, and the 
support of the ERC through an Advanced Grant. DCK and SMF were
supported by US NSF grant AST08-08133. VW acknowledges support of the
ERC through the starting grant SEDmorph. RJM acknowledges ERC funding via the award of a consolidator grant.
This work uses data from the following ESO programs: 60.A-9284,
181.A0717, 186.A-0898.
\end{acknowledgements}

\bibliographystyle{aa}

\end{document}